\documentclass[prb,twocolumn,showpacs,superscriptaddress,preprintnumbers,amsmath,amssymb]{revtex4}

\usepackage{graphicx}% Include figure files
\usepackage{dcolumn}% Align table columns on decimal point
\usepackage{bm}% bold math
\usepackage{color}

\newcommand{\be}{\begin{equation}}
\newcommand{\ee}{\end{equation}}
\newcommand{\ba}{\begin{eqnarray}}
\newcommand{\ea}{\end{eqnarray}}
\newcommand{\kv}{{\mathbf k}}
\newcommand{\kvv}{{\vec k}}
\newcommand{\xv}{{\mathbf x}}
\newcommand{\xvv}{{\vec x}}
\newcommand{\Real}{{\rm Re} \,}
\newcommand{\Imag}{{\rm Im} \,}
\newcommand{\eps}{\varepsilon}
\newcommand{\am}[1]{a^{\phantom \dagger}_{#1}}
\newcommand{\app}[1]{a^\dagger_{#1}}
\newcommand{\qv}{{\mathbf q}}
\newcommand{\qvv}{{\vec q}}
\newcommand{\Av}{{\mathbf A}}
\newcommand{\unit}{{\mathbf e}}
\newcommand{\g}[3]{g_{#1 #2}^{#3}}
\newcommand{\gconj}[3]{g_{#1 #2}^{#3 *}}
\newcommand{\f}[3]{f_{#1 #2, #3}}

\newcommand{\df}[3]{\dot{f}_{#1 #2, #3}}
\newcommand{\n}[2]{n_{#1 #2}}
\newcommand{\dn}[2]{\dot{n}_{#1 #2}}
\newcommand{\bmm}[1]{b^{\phantom \dagger}_{#1}}
\newcommand{\bp}[1]{b^\dagger_{#1}}
\newcommand{\V}[4]{V^{#1}_{#2 #3}(#4)}
\newcommand{\delpm}[2]{\delta(\eps_{#1 \kvv} - \eps_{#2 \kvv+\qvv} \pm \eps_{\rm LO})}
\newcommand{\delp}[2]{\delta(\eps_{#1 \kvv} - \eps_{#2 \kvv+\qvv} + \eps_{\rm LO})}
\newcommand{\delm}[2]{\delta(\eps_{#1 \kvv} - \eps_{#2 \kvv+\qvv} - \eps_{\rm LO})}
\newcommand{\fieldOpp}{\hat{\Psi}^\dagger(\xv,t)}
\newcommand{\fieldOpm}{\hat{\Psi}(\xv,t)}
\newcommand{\Hc}{{\rm H.c.}\,}
\newcommand{\Bosee}[1]{n_{#1}}
\newcommand{\Nph}[4]{N^{#1 #2}_{#3 #4}(\qv)}
\newcommand{\Nphconj}[4]{N^{#1 #2 *}_{#3 #4}(\qv)}
\newcommand{\Mimp}[4]{M^{#1 #2}_{#3 #4}(\qvv)}
\newcommand{\deltoo}[2]{\delta(\eps_{#1 \kvv} - \eps_{#2 \kvv+\qvv})}

\begin{document}

\title{Density-matrix theory of the optical dynamics and transport in quantum cascade structures: The role of coherence}

\author{C. Weber}
\email{carsten.weber@teorfys.lu.se}
\author{A. Wacker}
\affiliation{Mathematical Physics, Lund University, Box 118, 22100 Lund, Sweden}
\author{A. Knorr}
\affiliation{Institut f\"ur Theoretische Physik, Nichtlineare Optik und Quantenelektronik, Technische Universit\"at Berlin, Hardenbergstr. 36, 10623 Berlin}

\date{May 12, 2009, published as Phys. Rev. B 79, 165322 (2009)}

\begin{abstract}
The impact of coherence on the nonlinear optical response and stationary transport is studied in quantum cascade laser structures. Nonequilibrium effects such as pump-probe signals, the spatio-temporally resolved electron density evolution, and the subband population dynamics (Rabi flopping) as well as the stationary current characteristics are investigated within a microscopic density-matrix approach. Focusing on the stationary current and the recently observed gain oscillations, it is found that the inclusion of coherence leads to observable coherent effects in opposite parameter regimes regarding the relation between the level broadening and the tunnel coupling across the main injection barrier. This shows that coherence plays a complementary role in stationary transport and nonlinear optical dynamics in the sense that it leads to measurable effects in opposite regimes. For this reason, a fully coherent consideration of such nonequilibrium structures is necessary to describe the combined optical and transport properties.
\end{abstract}

\pacs{78.20.Bh,74.78.Fk,42.65.-k,73.63.-b}
\keywords{quantum cascade laser, heterostructure, nonlinear dynamics, pump-probe, transport}
\maketitle

\section{Introduction}

Quantum cascade lasers (QCLs)\cite{Faist:Science:94} are semiconductor heterostructure lasers employing the transitions between quantized intersubband levels in quantum well structures\cite{Elsaesser:PhysRep:99,Waldmuller::04a,Butscher:PhysRevB:05,Waldmuller:PhysRevB:06,Kira:PhysRevA:06} and act as a source of radiation in the terahertz (THz)/mid-infrared regime. The laser consists of multi-quantum well periods comprising the electron injector and the optically active region. These periods are repeated tens or even hundreds of times over the length of the structure.\cite{Gmachl:RepProgPhys:01} To drive the electrons through the sample, an external bias is applied. Scattering processes and optical recombination between the conduction subbands in the doped structure within as well as between periods lead to stationary electronic occupations out of equilibrium. %and transitions \todo{take out transitions?} 
While on the technological side, this light source can be used for spectroscopy in the fields of environmental detection or medicine,\cite{Gmachl:RepProgPhys:01,Lee:Science:07} it offers on a fundamental ground an interesting model system to study intersubband charge dynamics in a structure where the optical and the transport properties are closely interrelated.

%Semiconductor heterostructures were first investigated by Esaki and Tsu in 1970.\cite{Esaki:IBMJResDevelop:70} Since then, there has been a lot of experimental as well as theoretical work done in this field, in the interband\cite{Schilp:PhysRevB:94} as well as the intersubband regime.\cite{Waldmuller:PhysRevB:06,Butscher:PhysRevB:05,Kira:PhysRevA:06,Elsaesser:PhysRep:99,Waldmuller::04}
The first semiconductor heterostructure laser operating in the intersubband regime was realized by Faist {\it et al.} in 1994.\cite{Faist:Science:94}  Since then, many types of QCLs of different design have been built and optimized, see, e.g., Refs.~\onlinecite{Gmachl:RepProgPhys:01,Williams:NaturePhotonics:07} for an overview. The QCL has been the subject of extensive theoretical research. The stationary properties of QCLs were studied by a rate equation\cite{Harrison:ApplPhysLett:99,Indjin:JApplPhys:02} and a Boltzmann-type approach\cite{Iotti:ApplPhysLett:01,Callebaut:ApplPhysLett:03,Bonno:JApplPhys:05,Jirauschek:JApplPhys:07,Gao:JApplPhys:07,Jirauschek:PhysStatusSolidiC:08} as well as a quantum theory employing both nonequilibrium Green's functions\cite{Lee:PhysRevB:02,Banit:ApplPhysLett:05,Lee:PhysRevB:06,Kubis:PhysStatusSolidiC:08,Schmielau:PhysStatusSolidiB:09} and density-matrix theory.\cite{Iotti:PhysRevLett:01,Waldmueller:IEEEJQuantumElectron:06,Savic:PhysRevB:07} Here, the gain, the current-voltage characteristics, and the stationary charge distributions have been established. First results regarding the nonlinear optical properties such as optically induced subband population dynamics have been presented by us within a density-matrix theory.\cite{Weber:ApplPhysLett:06}

%Experimentally, coherent charge transport has been observed in optical pump-probe experiments,\cite{Eickemeyer:PhysRevLett:02,Eickemeyer:PhysicaB:02,Woerner:JPhys:CondensMatter:04,Darmo::08} showing a close relation between the optical and transport properties if the stationary nonequilibrium of the laser is perturbed on an ultrafast time scale.

One central result of these studies is that coherence can play an important role in the determination of the stationary current, requiring a fully microscopic theory of the current including non-diagonal density-matrix elements.\cite{Lee:PhysRevB:06,Callebaut:JApplPhys:05,Iotti:PhysRevB:05} Here, the often applied rate and Boltzmann equation approaches, based on Wannier-Stark hopping (WSH), fail.
%In the optical regime, the situation is not quite so clear.
Experimentally, indications of coherent charge transport have been obvserved in the oscillatory gain recovery in pump-probe experiments of mid-infrared QCLs\cite{Eickemeyer:PhysRevLett:02,Eickemeyer:PhysicaB:02,Woerner:JPhys:CondensMatter:04,Kuehn:ApplPhysLett:08} as well as recently in THz structures.\cite{Darmo::08} This oscillatory behavior has been attributed to resonant tunneling through the injection barrier of the laser. These observations suggest that coherent effects might also become visible in optics in the time regime beyond the light-matter interaction. However, different studies\cite{Choi:Phys:RevLett:08,Choi:ApplPhysLett:08} reveal a simple relaxation in the gain recovery, showing that not all samples exhibit this coherent effect.

Motivated by these investigations, this paper is focused on the role and importance of coherence in the interplay between ultrafast optical dynamics and stationary transport in quantum cascade structures. We investigate the regimes where coherence is of relevance in the combined optics-transport system. To this end, we present a fully microscopic theory describing the dephasing and tunneling  processes in a QCL structure. In order to systematically investigate the optical and transport regimes, we focus on two quantities describing the coherent and the incoherent evolution within the system: the tunnel coupling $2 \Omega$ between the two states across the main injection barrier and the level broadening $\Gamma$ of the states. We systematically vary the width of the main tunneling barrier in the QCL structure, thus establishing a relation between these two central quantities. Using this relation, we investigate the importance of coherence in the calculated signals: In the transport regime, we compare a rate equation (WSH) approach with a fully microscopic current theory to investigate the regime of validity of both models. In the optical regime, we consider pump-probe calculations, complemented by the spatio-temporally resolved electron density evolution and the subband population dynamics due to strong ultrafast excitation. Using these nonlinear results, aspects of the nature of the optically induced charge dynamics and the importance of coherence can be addressed.\cite{Iotti:PhysRevLett:01} 

Comparing the optical and transport results, we find that the influence of coherence is observable in opposite parameter regimes: the inclusion of coherence in stationary transport calculations becomes important for $2 \Omega \lesssim \Gamma$, where the coherence between subbands {\it limits} the current flowing through the structure, while in the nonlinear optical gain dynamics, the inclusion of coherence {\it drives} the observed oscillations in the gain recovery for $2 \Omega \gtrsim \Gamma$. Combining the two results, we thus find that it is necessary to consider a fully coherent theory in order to understand the combined optics-transport problem independent of the parameter range of operation.

%In Sec.~\ref{sect2}, we begin by introducing the theoretical formalism and the dynamical equations used in our calculations. In Sec.~\ref{sect3}, we consider the numerical results. We first establish a relation between the two central quantities of focus, the level broadening and the tunnel coupling across the main injection barrier of the QCL structure. Then, we focus on the stationary current calculations as well as the calculation of the nonlinear optical dynamics. Finally, a conclusion is given in Sec.~\ref{sect4}. The scattering equations are given in the  appendix.

\section{\label{sect2}Theory}

The QCL heterostructure is modeled as a multi-conduction subband system, each period comprising an injector and an active region. Within the effective mass and envelope function  approximations,\cite{Yu::99} the wave functions, assumed to be separable and infinitely extended in the quantum well plane, are given by
\be
\Psi_{i \kvv}(\xvv,z) = \frac{1}{\sqrt{A}} e^{i \kvv \xvv} \xi_i(z) u_{\kvv\approx0}(\xv),
\ee
where $A$ is the in-plane quantization area, $\xvv$ and $\kvv$ are the in-plane spatial and momentum vector, respectively, $\xi_i(z)$ is the envelope function in growth direction, and $u_{\kvv\approx0}(\xv)$ the Bloch function taken at the band edge. There are several natural possibilities to choose the wave functions $\xi_i(z)$ to describe the system. % (see Ref. \onlinecite{Wacker:PhysRep:02} for an overview of the advantages and disadvantages of the choice of basis functions).
\begin{figure}
\includegraphics[width=0.8\linewidth]{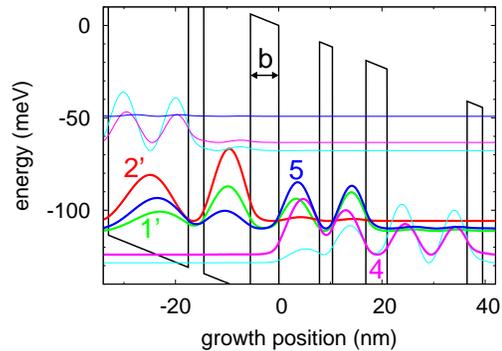} 
\caption{\label{WF}(Color online) Band structure of the THz QCL from Ref.~\onlinecite{Kumar:ApplPhysLett:04} for the injection barrier width $b$ = 5.5 nm under resonance condition. The subbands involved in the laser transition (4,1') as well as the injector subbands (5,2') are marked.}
\end{figure}
Since we consider the optical response of the system as well as the WSH model in the current calculation, we choose the Wannier-Stark (WS) basis which diagonalizes the heterostructure potential as well as the externally applied bias. This offers an intuitive physical interpretation of the optical transitions and the scattering between single-particle WS states as well as of the electron counting between approximate system eigenstates. It should be noted that, even though the physical observables are in principle independent of the choice of basis, different results are expected for different basis choices due to the necessary approximations applied in many-particle problems. In the following, the theory is applied to the THz QCL discussed in Ref.~\onlinecite{Kumar:ApplPhysLett:04}. The band structure of this QCL in the original design is shown in Fig.~\ref{WF} with the important subbands marked. This structure is considered throughout this paper, considering different injection barrier widths $b$. The parameters used in the calculations are found in Table~\ref{QCLstructure2}.

In order to describe the ultrafast nonlinear optical and stationary transport properties of the laser, it is preferable to consider a fully microscopic treatment of the scattering mechanisms in the structure. In this work, we consider the interaction of the electronic system with longitudinal optical (LO) phonons via the polar (Fr\"ohlich) coupling as well as with the ionized doping centers in the laser. Even though Coulomb scattering can be treated in the same way,\cite{Waldmuller:PhysRevB:04,Shih:PhysRevB:05} we do not consider this interaction here. For the relatively lowly doped THz laser considered here, it was shown that scattering with impurities typically dominates over Coulomb scattering.\cite{Callebaut:ApplPhysLett:04} Acoustic phonons are excluded since they induce only very long scattering times in quantum wells and act mainly as a low-energy intrasubband thermalization mechanism.\cite{Nelander:PhysStatusSolidiC:09} Interface roughness scattering is excluded due to the difficulty to quantify it microscopically, while alloy scattering is small in structures with a binary material for the wells (GaAs) and therefore neglected.\cite{Vasanelli:ApplPhysLett:06} The included scattering mechanisms are treated in a density-matrix correlation expansion approach within a second order Born-Markov approximation.\cite{Fricke:AnnPhys:96,Rossi:RevModPhys:02} We do not include renormalizations of the subband energies due to scattering in the form of principal values (see Sec.~\ref{secIIB3} for a discussion of this approximation); it is assumed that these are small and approximately constant for the different transitions, leading only to an absolute energy shift without physical consequences (see Ref.~\onlinecite{Butscher:PhysRevLett:06} for a discussion of scattering-induced energy renormalizations in quantum wells).
\begin{table}
\caption{\label{QCLstructure2}Structural and material parameters used in the calculations.}
\begin{ruledtabular}
\begin{tabular}{l c c}
material system	&					&GaAs/Ga$_{0.15}$Al$_{0.85}$As\\
doping density &$n_{\rm 2d}$				&2.945$\times 10^{10}$ cm$^{-2}$\\
well electron mass &$m^*$					&0.067 m$_0$\\
LO phonon energy &$\hbar \omega_{\rm LO}$		&36.7 meV\\
high-frequency permittivity &$\eps_\infty$	&10.9\\
static permittivity &$\eps_s$		&12.9
\end{tabular}
\end{ruledtabular}
\end{table}

\subsection{Hamiltonian}

To derive the dynamical equations, we divide the Hamiltonian of the system into three parts:
\be
H = H_0 + H_{\rm el-light} + H_{\rm scatt}.
\ee
The first part,
\be
H_0 = H_{\rm SL} + H_{\eps} + H_{\rm 0,ph} = \sum\limits_{i \kvv} \eps_{i \kvv} \app{i \kvv} \am{i \kvv} + \sum\limits_\qv \hbar \omega_\qv b_\qv^\dagger b_\qv^{\phantom \dagger},
\ee
describes the kinetics of the electrons in the heterostructure potential $H_{\rm SL}$ as well as the externally applied bias $H_{\eps}$ and the kinetics of the phonons $H_{\rm 0,ph}$. The electronic part is diagonal due to the choice of basis. $a_{i \kvv}^\dagger$ ($a_{i \kvv}^{\phantom \dagger}$) denotes the creation (annihilation) operator of an electron in subband $i$ with quasi-momentum $\kvv$ and energy $\eps_{i \kvv} = \eps_{i} + \hbar^2 \kvv^2 / (2 m_i)$ and $b_{\qv}^\dagger$ ($b_{\qv}^{\phantom \dagger}$) the creation (annihilation) operator of an LO phonon with three-dimensional quasi-momentum $\qv$ and energy $\hbar \omega_\qv \equiv \hbar \omega_{\rm LO}$. Subband nonparabolicity is neglected, and the subband effective masses are assumed to be constant, $m_i \equiv m^*$, where $m^*$ is the effective mass of the well material.

$H_{\rm el-light}$ describes the interaction of the system with a coherent classical light source. The polarization of the light field is chosen in the direction of the dipole moment, i.e. in the growth ($z$-) direction, and the field is assumed to be a spatially homogeneous Gaussian pulse,
\be
\Av(t) = A(t) \unit_z, A(t) = A_0 \exp \{ -( t/\tau )^2 \} \cos(\omega_L t),
\ee
with the laser frequency $\omega_L$ and Gaussian pulse duration $\tau$. The pulse area is defined via the envelope of the pulse.\cite{Allen::87} Under the assumption of a homogeneous excitation, the Hamiltonian is momentum-diagonal and reads
\be
H_{\rm el-light} = e A(t) \sum\limits_{i j} \sum\limits_\kvv M_{ij} \app{i \kvv} \am{j \kvv}\label{HamElLight}
\ee
with the elementary charge $e>0$, the coupling elements $M_{ij} = \frac{1}{2} \langle i| [\hat{p}_z/m(z) + e A(t)/(2 m(z))] + {\rm H.c.} |j\rangle$, and the momentum operator $\hat{p}_z = (\hbar/i) \partial_z$ with the space-dependent effective mass $m(z)$.% The $A^2$ term of the light-matter interaction has been shown to be weak even in the cases of strong excitation considered in this work and is thus neglected.

Finally, $H_{\rm scatt}$ describes the scattering processes: electron-LO phonon interaction $H_{\rm el-ph}$ as well as scattering with ionized doping centers $H_{\rm el-imp}$:

\subsubsection{Electron-LO phonon interaction}

The electron-LO phonon coupling Hamiltonian is given by
\be
H_{\rm el-ph} = \sum\limits_{i j} \sum\limits_{\kvv,\qv} \g{i}{j}{\qv} \app{i \kvv} \bmm{\qv} \am{j \kvv-\qvv} + \gconj{i}{j}{\qv} \app{j \kvv-\qvv} \bp{\qv} \am{i \kvv},
\ee
where $\qvv$ is the in-plane projection of $\qv$. The coupling matrix element is given by the Fr\"ohlich coupling
\be 
\g{i}{j}{\qv} = -i \left[ \frac{e^2 \hbar \omega_{\rm LO}}{2 \eps_0 V} \left( \frac{1}{\eps_\infty} - \frac{1}{\eps_s} \right) \right]^{1/2} \frac{\unit(\qv) \cdot \qv}{q^2} \langle\xi_i|e^{i q_\perp z}|\xi_j\rangle.
\ee
Here, $\eps_s$ and $\eps_\infty$ are the static and high-frequency permittivity, respectively, $V$ is the quantization volume, $\unit(\qv)$ is the displacement unit vector, and $q_\perp$ the projection of the momentum $\qv$ in growth direction.

\subsubsection{Interaction with ionized doping centers}

Since the quantum cascade structure is doped, it is necessary to take into account the interaction of the electrons with the ionized doping centers. Typically, either a barrier or a well in the QCL is doped, making it necessary to distribute the ions in this layer. Here, we treat the interaction following Ref.~\onlinecite{Banit:ApplPhysLett:05}, where the dopant density is distributed on several $\delta$-sheets located at $z_l$ in the doped barrier/well. The ionized doping centers are considered as classical scattering centers. The Hamiltonian for the interaction is given by
\be
H_{\rm el-imp} = \sum\limits_{i j} \sum\limits_{\kvv,\qvv} \sum\limits_l V^l_{i j}(\qvv) \app{i \kvv+\qvv} \am{j \kvv}
\ee
with the screened electron-impurity interaction potential
\be
\V{l}{i}{j}{\qvv} = \frac{1}{A} \frac{-e^2}{2 \eps_0 \eps_s \sqrt{\qvv^2 + \lambda^2}} \langle\xi_i|e^{-i \qvv \xvv_l} e^{-\sqrt{\qvv^2 + \lambda^2} |z-z_l|}|\xi_j\rangle.
\ee
Here, $\lambda$ is the screening constant and $l$ labels the position of the randomly distributed individual ion $\delta$-distributions in the growth direction. For the screening, we use the static limit of the Lindhard formula for a homogeneous electron gas.\cite{Haug::04}

\subsection{Dynamical equations}

\subsubsection{Equation structure and approximations}

We derive the dynamical equations using a correlation expansion within a density-matrix approach in second order Born-Markov approximation, applying a bath approximation for the LO phonons.\cite{Fricke:AnnPhys:96,Rossi:RevModPhys:02} This leads to equations of motion for the microscopic polarizations $\f{i}{j}{\kvv} = \langle \app{i \kvv} \am{j \kvv} \rangle (i \neq j)$ and the subband occupations $\n{i}{\kvv} = \langle \app{i \kvv} \am{i \kvv} \rangle$. On the level of the phonon-(impurity)-assisted coherences $\langle a^\dagger_{i \kvv} b_{\bf q}^{(\dagger)} a^{\phantom \dagger}_{j \kvv'} \rangle$ ($\langle a^\dagger_{i \kvv} V^l_{m n}(\qvv) a^{\phantom \dagger}_{j \kvv'} \rangle$), respectively, we neglect the interaction with the optical field since these terms are of higher order in the coupling.\cite{Schilp:PhysRevB:94} We assume the system to be homogeneous in the plane perpendicular to the growth direction $z$; then, we can restrict to a density matrix diagonal in the in-plane momentum: $\langle\app{i \kvv} \am{j \kvv'}\rangle = \f{i}{j}{\kvv} \delta_{\kvv \kvv'}$.

The general dynamical equations are found in the appendix. While for the interaction with the doping centers, the complete equation structure is considered, only the terms linear in the density matrix are taken along for the electron-phonon interaction due to numerical reasons. This approximation is justified if we assume that we are working in the regime of nondegenerate electron gases, where $\n{i}{\kvv} \ll 1$ which is typically fulfilled in these QCL structures.

We should stress at this point that, due to the spatial extension of the Wannier-Stark wave functions, it is essential to consider the whole set of matrix elements in the calculations. Testbed calculations considered only certain sets of matrix elements, e.g., we restricted to the diagonal/nondiagonal scattering terms $\g{i}{j}{\qv} \gconj{i}{j}{\qv}$, considered typically in nonbiased (equilibrium) quantum well systems,\cite{Waldmuller:PhysStatusSolidiB:03,Waldmuller:PhysRevB:04,Shih:PhysRevB:05,Waldmuller:PhysRevB:06} or alternatively to terms where the overlap of the wave functions is the main argument, i.e. terms such as $\g{i}{i}{\qv} \gconj{i}{j}{\qv}$ along with the above set of terms. Both versions lead to nonstable results in some parameter regimes where the coherences are important. It is thus essential to consider the full set of matrix elements when considering the properties of the QCL at resonance condition and for large injection barrier widths, i.e. small tunnel coupling across the injection barrier.

When considering the WSH model, the relaxation process are restricted to Boltzmann scattering given by
\be \label{WSHeqs}
\dot{n}_{i \kvv}^{\rm (0)} = -\Gamma^{\rm out}_{i \kvv} n_{i \kvv} + \Gamma^{\rm in}_{i \kvv} (1 - n_{i \kvv}).
\ee
The in- and out-scattering rates $\Gamma^{\rm in}_{i \kvv}$ and $\Gamma^{\rm out}_{i \kvv}$ are functions of the subband occupations $\n{l}{\kvv'}$ and are \
given for both the electron-phonon and electron-impurity interaction in the appendix.\\

The system temperature enters the calculations through the scattering rates which include the phonon distribution, assumed to be given by an equilibrium Bose-Einstein distribution $n_{\bf q} = [\exp(\hbar \omega_{\bf q}/k_B T)-1]^{-1}$, and the screening parameter $\lambda$ for the interaction of the electrons with the doping centers, see also Ref.~\onlinecite{Nelander:ApplPhysLett:08}. Here, the approximation is made that the lattice and the electronic temperature are the same.

\subsubsection{Periodic boundary conditions}

In order to describe the extension of the periodically coupled structure, it is necessary to apply appropriate boundary conditions in the growth direction. Since the inclusion of coherence between states of different periods is important to describe the nonlinear optical dynamics due to the spatial extension of the wave functions,\cite{Weber:ApplPhysLett:06} we consider a nearest neighbor approach where the coherence and wave function overlap between two adjacent periods $(n,n')$ is taken into account. We checked this approximation by showing that $|\f{i n}{j n'}{\kvv}|$, $\g{i n}{j n'}{\qv},\V{l}{i n}{j n'}{\qvv} \approx 0$ for $|n-n'|\geq2$, where $\f{i n}{j n'}{\kvv}$ ($\g{i n}{j n'}{\qv},\V{l}{i n}{j n'}{\qvv}$) denotes the coherence (coupling element) between the states $|i\rangle$ and $|j\rangle$ in the periods $n$ and $n'$, respectively. We apply translational invariance of the density matrix between periods: \footnote{Note that this result depends on the gauge for the exciting field which is used. In the case of the Lorenz gauge, it is necessary to take into account a phase factor when transforming the density matrix between different periods.}
\be
\f{i n}{j n'}{\kvv} = \f{i (n+1)}{j (n'+1)}{\kvv}, \qquad n_{i n, \kvv} = n_{i (n+1), \kvv}.
\ee
For the wave functions and the band edge energies $\eps_i$, we apply a coordinate shift and a bias drop, respectively,
\be
\xi_{i n}(z) = \xi_{i (n+1)}(z+L_{\rm per}), \qquad \eps_{i n} = \eps_{i (n+1)} - eEL_{\rm per},
\ee
where $L_{\rm per}$ is the period length and $E$ the applied electric field. In the following, the index $i \equiv (i,n)$ is taken as a composite index.

\subsubsection{Shortcomings of the model} \label{secIIB3}

\begin{figure}
\includegraphics[width=0.49\linewidth]{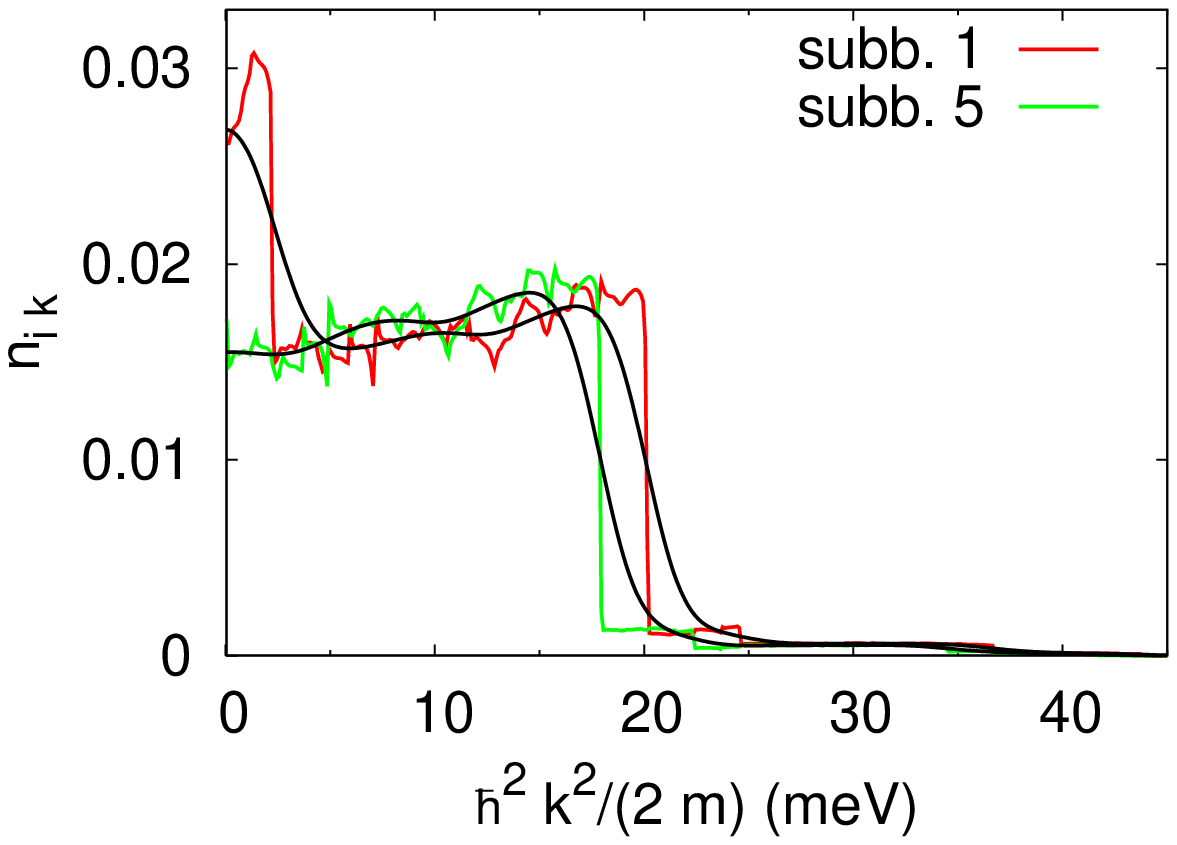}
\includegraphics[width=0.49\linewidth]{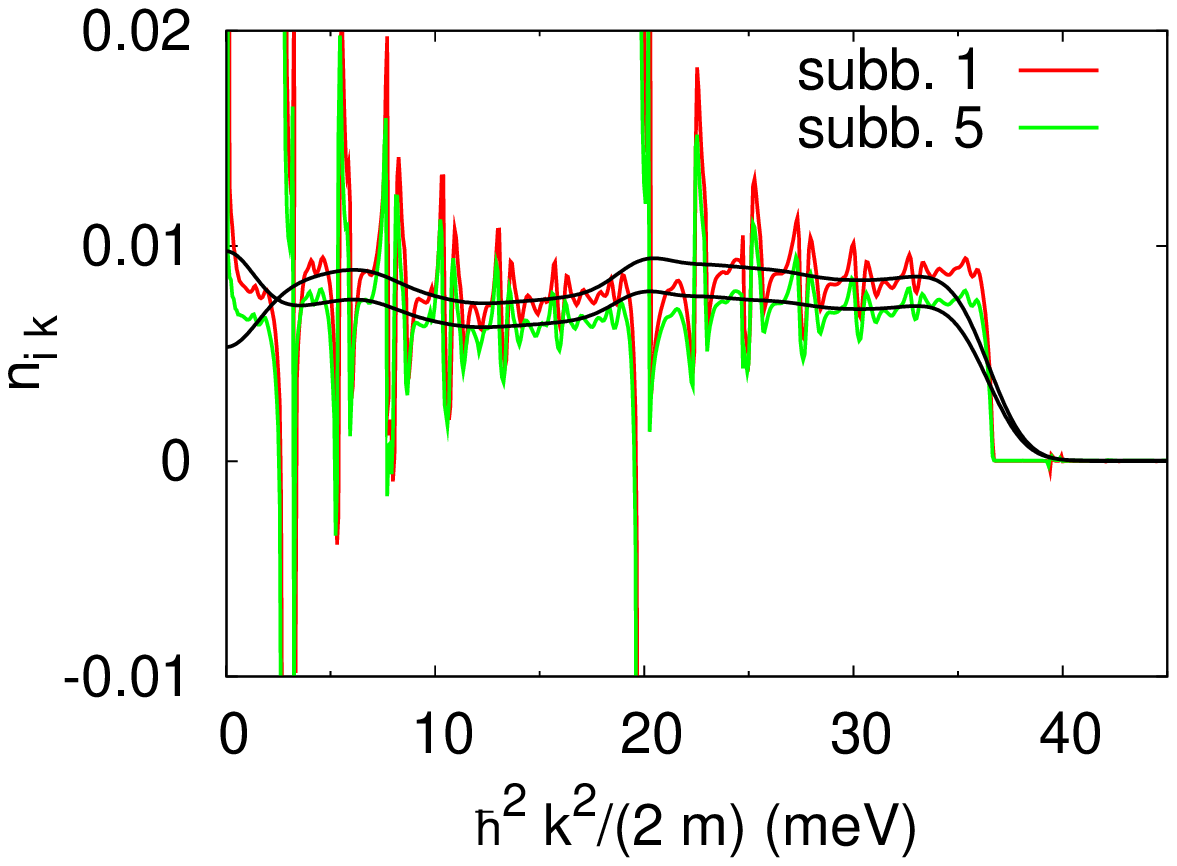}
\caption{\label{distribs}(Color online) Gaussian smoothening of the subband occupations $\n{i}{\kv}$ for the two main subbands (injector $5$, upper laser state $1$) at $T$ = 10 K: (left) $b$ = 4.5 nm, (right) $b$ = 11.0 nm.% (right) Distribution functions of the two resonant subbands for the barrier widths $b$ = 4.5 nm (thin lines) and $b$ = 13.0 nm (thick lines), showing a heating up of the injector and upper laser state which leads to higher electron-phonon scattering rates from these two states.
}
\end{figure}

Even taking along the whole set of equations within the approximations discussed above, we partially encounter negative occupations $\n{i}{\kv}$ (see Fig.~\ref{distribs}) which is known from the treatment of Redfield-type equations such as considered here.\cite{Breuer::02,Whitney:JPhysA:MathTheor:08,Schaller:PhysRevA:08} However, this typically does not lead to unphysical results if we focus on averaged observables. For all dynamical calculations, we obtain physical values for the current density, the total electron densities in each subband, and the gain spectrum. However, the broadening in Eq.~(\ref{broadening}) may become negative for regimes where coherences play an important role due to the negative occupations. This may be due to the fact that (i) the scattering rates as such are not observables and (ii) the simple formula for the broadening used here, which is typically applied in the literature, does not take into account stationary coherence. In order to guarantee strictly positive values for the averaged scattering rates to calculate the broadening in Eq.~(\ref{broadening}), which we can use as a measure of the lifetime of the Wannier-Stark states, we perform a Gaussian smoothening of the stationary distributions and use these in the calculations of the broadening $\Gamma$. Figure~\ref{distribs} shows an example of this smoothening for two different barrier widths: While for small barrier widths and low temperatures, the negative occupations are negligible, and thus the corresponding averaged scattering rates are always positive, the smoothening becomes important for large barrier widths where the negative occupations can lead to negative values. Thus, the values obtained for the broadening $\Gamma$ should be viewed as approximative values. Since we are concerned with qualitative results only, this is justified. For the cases where the negative occupations are negligible, and thus the averaged scattering rates strictly positive, the smoothening induces changes of the broadening of $2-3\%$. We would like to stress again, however, that all dynamical calculations and all results except for Fig.~\ref{omegaGamma} are performed without artificial modifications of the calculated data, such as any kind of smoothening. In addition, it should be noted that the calculations here are performed at the point of ``maximal coherence'' with respect to the tunnel coupling across the injection barrier, i.e. at resonance (cf. Sec.~\ref{sect3.1}). Away from resonance, the influence of coherence, and the correspondingly related problems such as non-positivity, are strongly reduced.

The reason for the quite large negative values for the occupations partially encountered may be explained by the fact that the system under consideration is a very complicated one: for the nearest neighbor coupling considered in the calculations, we have 35 independent coherences, and thus there is a very strong interplay between the different frequencies of the coherences which can destroy the strict positivity of the occupations. For small barrier widths and low temperatures, the general dynamical equations (\ref{elPhEqs}) and (\ref{elDotEqs}) reduce to the Boltzmann dynamics given in Eq.~(\ref{WSHeqs}). Here, any negative occupations observed in our full calculations are negligible ($\n{i}{\kv} \gtrsim -10^{-8}$) [see, e.g.,  Fig.~\ref{distribs}(left)]. In the case of the Boltzmann scattering dynamics, i.e. including only the diagonal elements of the density matrix, the stationary occupations remain strictly positive for all barrier widths and temperatures.

The specific oscillation energies of the peaks can be explained by the energetic structure of the system: the peaks correspond to different combinations of the bias drop per period $eFL_{\rm per}$ and the LO phonon energy $E_{\rm LO}$. In Fig.~\ref{distribs}(right), e.g., the large peaks correspond to the energy $eFL_{\rm per} - E_{\rm LO} = 56.40$ meV $- 36.7$ meV = $19.7$ meV, while the smaller peak positions are given by $2 eFL_{\rm per} - 3 E_{\rm LO} = 2.7$ meV. The latter are weaker, since they represent a higher order process of iterative transport and scattering. For the two subbands, the peaks are shifted by $\Delta E \approx 0.1$ meV, corresponding to the tunnel splitting between the two states, $E_{1'} - E_5 \approx 0.1$ meV.

As stated above, there is a strong interplay between different coherences of sharp energy. For $T$ = 50 K and a certain barrier width regime, we need to incorporate additional damping in order to obtain convergence. In the calculations, we do this by adding a phenomenological, low-energy intrasubband scattering mechanism with an energy which is uncommensurable with the LO phonon energy ($E \approx$ 0.5 meV) and use the LO phonon coupling element with an increased strength. We should note that this damping is of a purely phenomenological character and not a physical scattering mechanism as implemented here. The concerned barrier widths are marked in Figs.~\ref{omegaGamma} and \ref{currBarr}. The problem with convergence may be a consequence of the combination of neglecting low-energy scattering channels and applying the Born-Markov approximation within the density-matrix theory. Another explanation of this convergence problem is that scattering-induced energy renormalizations are neglected at this order of the perturbation expansion. This is an approximation often applied within the Born-Markov approximation.\cite{Waldmuller:PhysRevB:04,Savic:PhysRevB:07} Since the renormalizations, like the scattering rates, are temperature-dependent, this could explain why the convergence problems only appear for certain temperatures. On the other hand, since the resonances depend on the specific energy structure, and thus on the barrier width $b$, problems are expected to appear only for certain barrier widths, as witnessed in the results.

Including an artificial damping as discussed above, we find that the value of the stationary current is approximately independent of the scattering strength. This has already been observed in earlier studies of the QCL (Refs.~\onlinecite{Jirauschek:JApplPhys:07,Nelander:PhysStatusSolidiC:09}) and attributed to the fact that a low-energy dissipative scattering mechanisms such as LA phonons leads mainly to a thermalization within the subbands, but has no strong effect on the actual value of the stationary current. We adopt the validity of this statement here and take the obtained values of the current as physical results. On the other hand, it is clear that the inclusion of additional large damping leads to a strong overestimation of the total broadening as calculated in Eq.~(\ref{broadening}). Since the relation between the broadening $\Gamma$ and the tunnel coupling $2 \Omega$ is used to discriminate qualitatively different regimes, the inclusion of additional scattering mechanisms for the concerned barrier widths (which typically leads to a larger broadening) does not lead to qualitatively new regimes of interest (see Fig.~\ref{omegaGamma}). It is for this reason that one can safely disregard these values in the discussion of the broadening and the tunnel coupling as is done in the following.

\section{\label{sect3}Numerical Results}

In this section, we apply the theory of Sec.~\ref{sect2} to the THz laser from Ref.~\onlinecite{Kumar:ApplPhysLett:04}. This structures has been investigated with respect to its transport in the stationary laser regime as well as its gain properties.\cite{Banit:ApplPhysLett:05} In the regime of ultrafast nonlinear optics, Rabi oscillations were considered recently.\cite{Weber:ApplPhysLett:06} 

First, the systematic setup is discussed, along with the calculation of the central quantities used for the further discussion of the tranport and optical properties. The stationary nonequilibrium due to the different scattering mechanisms is determined. This is necessary for the determination of the stationary value of the current, but also as a starting point for the consideration of the optical response. Then, the WSH approach is compared with a full current calculation to investigate the importance of the inclusion of coherence in the stationary transport. Finally, the nonlinear optical response of the structure is considered, focusing on pump-probe signals as well as the spatio-temporally resolved electron density and the optically induced population dynamics, to study the role of coherence in the optical regime.

\subsection{\label{sect3.1}Systematic setup and stationary state}

\begin{figure}
\includegraphics[width=0.54\linewidth]{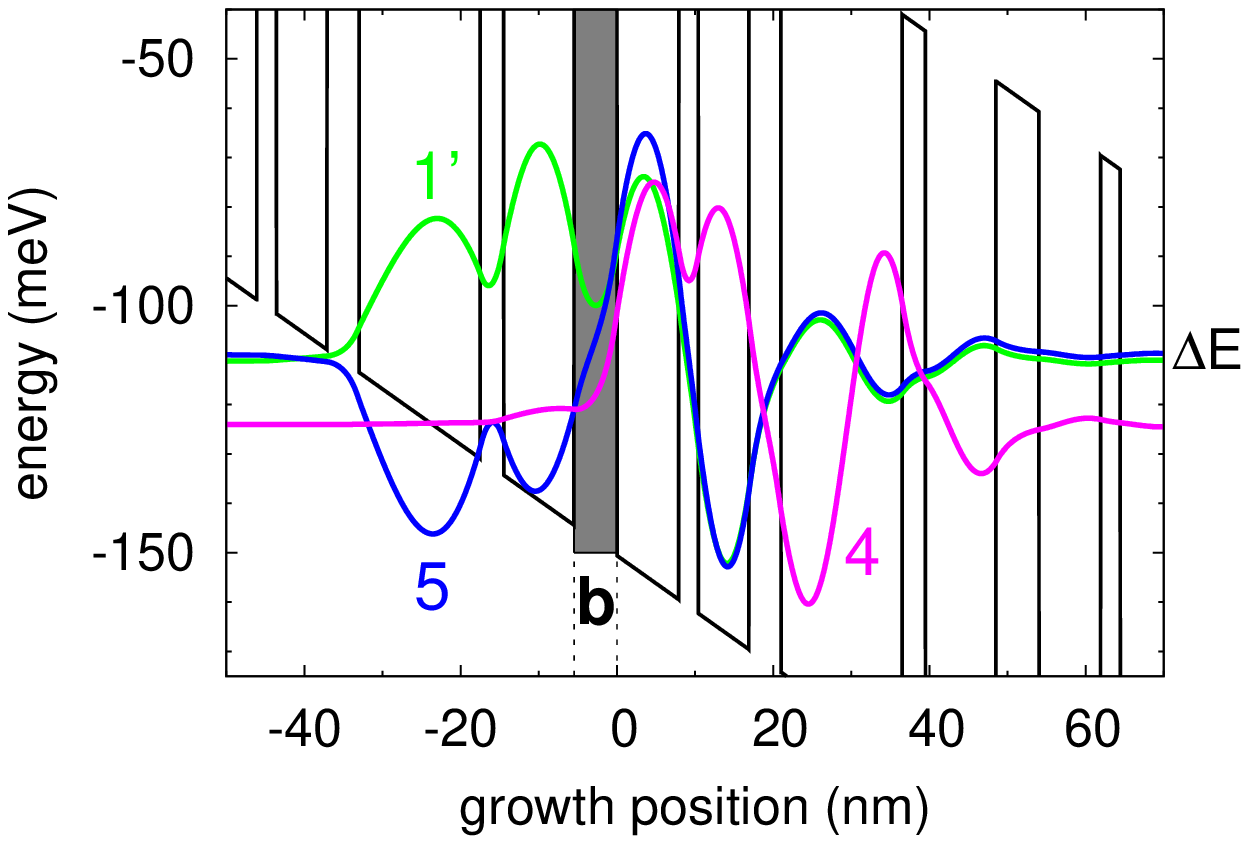}
\includegraphics[width=0.44\linewidth]{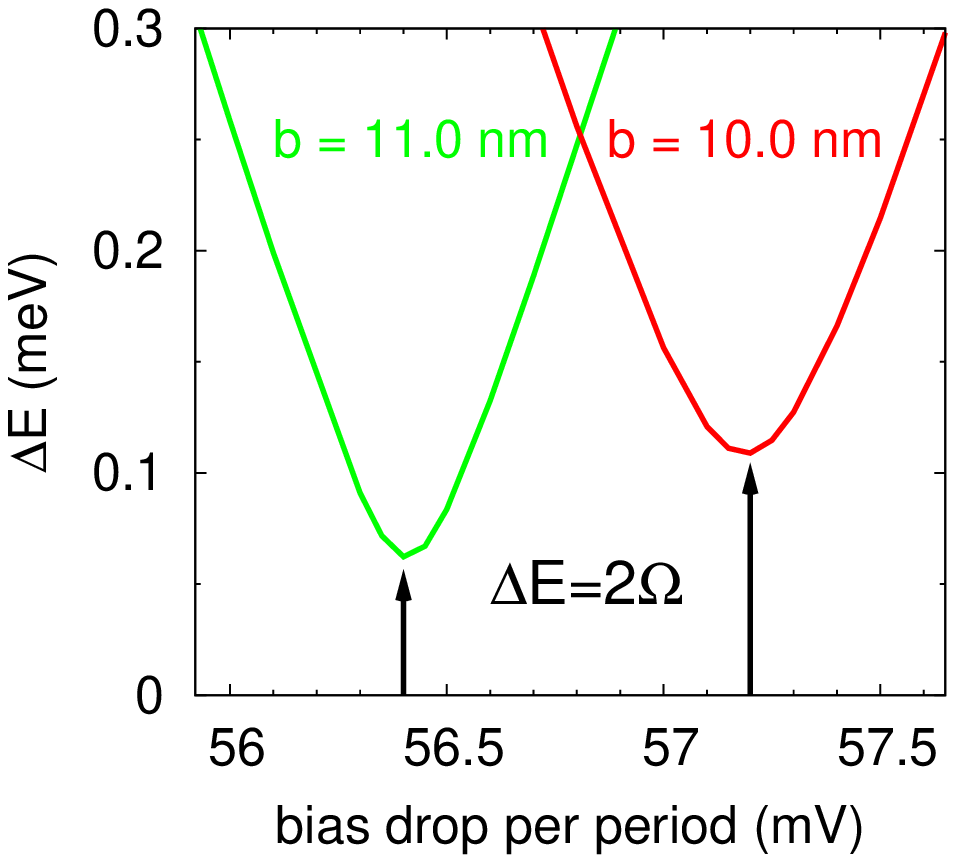}
\caption{\label{anticross}(Color online) (left) Injector (5) and upper laser state (1') showing an anti-crossing at the main tunneling barrier at exact resonance of the corresponding Wannier states. The lower laser state (4) is also shown. (right) Tunnel splitting energy $\Delta E = |E_5 - E_{1'}|$ for varying bias and different barrier widths $b$ determining the resonance bias at the center of the anti-crossing and the tunnel coupling 
$2 \Omega \approx \Delta E$.}
\end{figure}

In order to systematically investigate the importance of coherence, we vary the width of the main tunneling barrier connecting the injector and the active region of the laser. When resonant tunneling between the injector and the upper laser state is small, the WS states are approximately localized in the injector and in the active region. For a certain applied bias, the WS states become delocalized across the injection barrier and an anticrossing occurs, where the injector and the upper laser level form a pair of binding/antibinding states. The system is assumed to be {\it in resonance} at the center of the anticrossing, i.e. for the bias at which the splitting energy between the two respective levels has a minimum. Here, the level splitting $\Delta E$ equals twice the tunnel coupling $2 \Omega$ between the localized injector and the upper laser state, see Fig.~\ref{anticross}.

\begin{figure}
\includegraphics[width=0.49\linewidth]{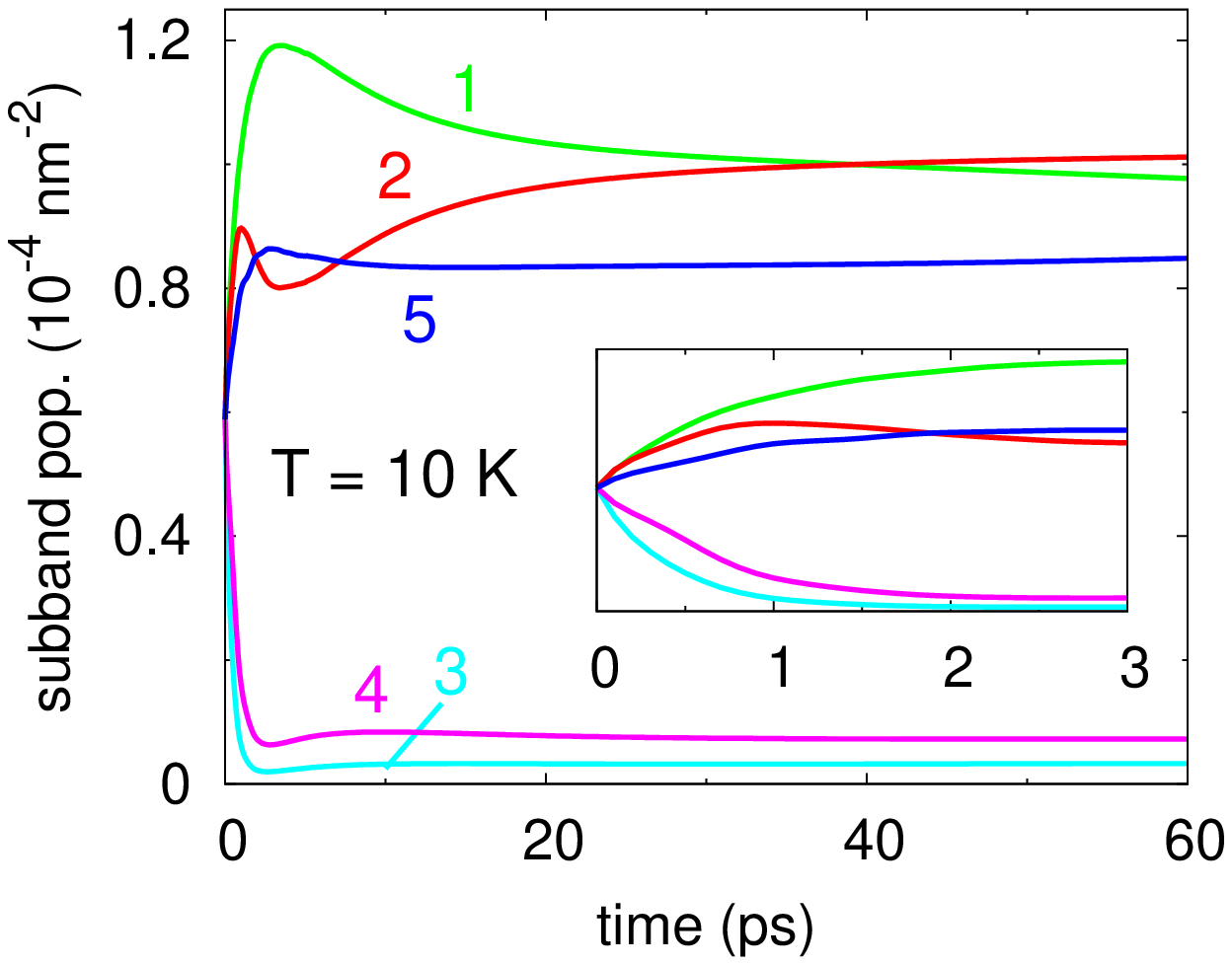}
\includegraphics[width=0.49\linewidth]{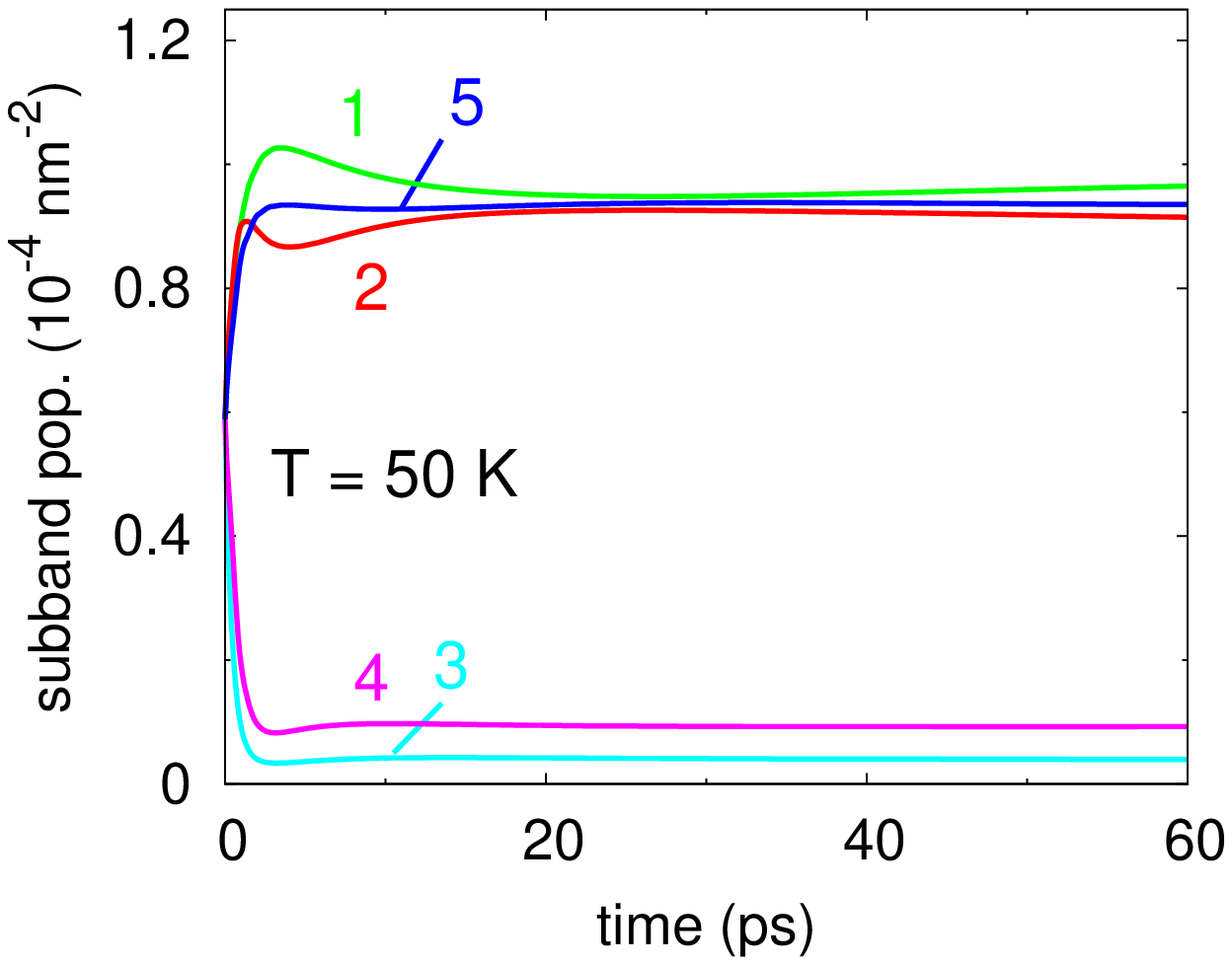}
\caption{\label{relax}(Color online) Population relaxation dynamics using the full set of dynamical equations for a barrier width of $b$ = 5.5 nm at $T$=10 K, 50 K, determining the stationary nonequilibrium of the system. Inset: Short-time evolution of the subband populations.}
\end{figure}

The QCL as a multi-subband nonequilibrium system is a very complex structure, and thus a straight-forward analytical determination of the stationary population and coherence distributions due to scattering is not possible. We thus numerically determine the initial conditions to be stationary solutions for $\f{i}{j}{\kvv}$ and $\n{i}{\kvv}$ without the optical field by solving all scattering contributions from an arbitrary state of fixed total population until a steady state is reached. Figure \ref{relax} shows the process of determining the initial distributions by starting from subband populations $n_i = 2/A \sum_\kvv n_{i \kvv}$ given by single-subband Fermi distributions with equal populations in each subband.
An approximate stationarity is reached on a time scale of a few picoseconds, while long time intraband redistributions can last up to several nanoseconds (not shown). For small barrier widths, i.e. large tunnel coupling $\Omega$, the relaxation is described well by a Boltzmann relaxation $\dn{i}{\kvv}^{(0)}$. For larger barrier widths, the influence of the coherence on the stationary populations cannot be neglected any longer and becomes important which is discussed in detail in the next section.

Having determined the stationary distributions, we can calculate the scattering induced broadening of the states. %As for the determination of the tunnel coupling, 
The level broadening of the considered states is approximately given by the mean of the two tunnel-split Wannier-Stark states, $\Gamma = (\Gamma_{1'} + \Gamma_5)/2$, where the broadening is determined by the stationary Boltzmann out-scattering rate,
\be\label{broadening}
\Gamma_i = 
%\frac{\sum_\kvv \n{i}{\kvv}^{(0)} \Gamma^{\rm out}_{i \kvv} \{ \n{}{}^{(0)} \} }{\sum_\kvv \n{i}{\kvv}^{(0)}} \equiv 
\frac{\frac{2}{A} \sum_\kvv \n{i}{\kvv}^{(0)} \; \Gamma^{\rm out}_{i \kvv} \{ \n{l}{\kvv}^{(0)} \} }{\n{i}{}^{(0)}},
\ee
% \be
% \Gamma_i = \frac{\frac{2}{A} \sum\limits_\kvv \left( \Gamma^{\rm in}_{i \kvv} \{ \n{}{}^{(0)} \} + \Gamma^{\rm out}_{i \kvv} \{ \n{}{}^{(0)} \} \right) \n{i}{\kvv}^{(0)}}{\n{i}{}^{(0)}},
% \ee
for the stationary occupations $\n{l}{\kvv}^{(0)}$.
%The broadening is shown in Fig.~\ref{gamma} for two different temperatures. As mentioned in Sec.~\ref{sect2}, a Gaussian smoothening of the stationary occupations has been performed for the calculation of the scattering rates in order to circumvent the problem with negative occupations obtained due to the structure of the dynamical equations. An example of the smoothening for two subbands is shown in Fig.~\ref{distribs}(left).
% \begin{figure}[h]
% \begin{centering}
% \includegraphics[width=0.49\linewidth]{pics/gamma_10K.eps}
% \includegraphics[width=0.49\linewidth]{pics/gamma_50K.eps}
% \end{centering}
% \caption{(Color online) Mean broadening $\Gamma = (\Gamma_{5} + \Gamma_{1'})/2$ of the injector and the upper laser state for different barrier widths $b$ and at $T$ = 10 K, 50 K for the different scattering mechanisms.}
% \label{gamma}
% \end{figure}
At these low temperatures, the broadening is typically dominated by the impurity scattering, which is strongly temperature dependent due to the screening length\cite{Nelander:ApplPhysLett:08} and thus increases strongly from 10 K to 50 K, while the interaction with phonons is essentially given by spontaneous emission.
%At larger barrier widths, the increased population in the injector $5$ and upper laser state $1'$ is accompanied by a heating up [see Fig.~\ref{distribs}(right)], and thus scattering via phonon emission both intraband and interband into the other resonant state as well as into the lower laser state and the extractor increases strongly due to the broader distribution functions. The extent of this heating up would probably be less in a realistic device due to intraband LA phonon scattering which would thermalize the distribution functions and thus decrease the stationary occupations for larger energies and with that the scattering rates from these two subbands. The scattering rates for the impurities remain roughly constant.
% \begin{figure}[h]
% \begin{centering}
% \includegraphics[width=0.49\linewidth]{pics/distribSmooth.eps}
% \includegraphics[width=0.49\linewidth]{pics/distribHeating.eps}
% \end{centering}
% \caption{(Color online) (left) Gaussian smoothening of the distribution functions. (right) Distribution functions of the two resonant subbands for the barrier widths $b$ = 4.5 nm (thin lines) and $b$ = 13.0 nm (thick lines), showing a heating up of the injector and upper laser state which leads to higher electron-phonon scattering rates from these two states.}
% \label{distribs}
% \end{figure}

\begin{figure}
\includegraphics[width=0.9\linewidth]{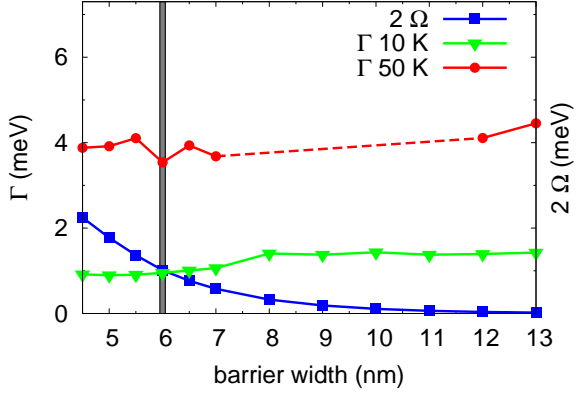}
\caption[]{\label{omegaGamma}(Color online) Tunnel coupling $2 \Omega$ and mean broadening $\Gamma = (\Gamma_{5} + \Gamma_{1'})/2$ of the injector and the upper laser state for different barrier widths $b$ and at $T$ = 10 K, 50 K.\footnotemark[1]}
\end{figure}

%Having calculated these two central quantities, we now have a relation between the barrier width and the corresponding tunnel coupling between and broadening of the two considered states. This is shown in Fig.~\ref{omegaGamma} for the two  temperatures.
Figure~\ref{omegaGamma} shows the calculated values of $\Omega$ and $\Gamma$ as a function of the barrier width $b$. As is expected, the broadening remains roughly constant for varying barrier widths, since the scattering rates are not strongly influenced by the width, while the tunnel coupling across the barrier approaches zero for $b \rightarrow \infty$. For $T$ = 10 K, the two energies are roughly the same for $b \approx 6.0$. This is the regime which is physically interesting, and thus barrier widths slightly larger and less than this value are the focus in the following investigations. For $T$ = 50 K, the broadening of the states is much larger then the tunnel coupling for all considered barrier widths, and thus no transition between different regimes is expected.
\renewcommand{\thefootnote}{\alph{footnote}}
\footnotetext[1]{For the barrier widths marked by dashed lines (open circle in the current density), the inclusion of an additional low-energy scattering mechanism is necessary to achieve convergence of the current (see the discussion in the text). The value of the current density obtained in this way is approximately independent of the strength of the additional coupling when determined from the transient, in accordance with earlier results.\cite{Jirauschek:JApplPhys:07,Nelander:PhysStatusSolidiC:09} The resulting, unphysically large total broadening $\Gamma$ does not lead to qualitatively new regimes of interest and thus not considered further.}
\renewcommand{\thefootnote}{\number{footnote}}

\subsection{Stationary results: Current calculations}\label{secIIIB}

The stationary distributions far from equilibrium determined in Sec.~\ref{sect3.1} are the starting point for the optical dynamics discussed later in this chapter. Here, we consider the accompanying evolution to the steady state value of the current flowing through the structure. In order to investigate the influence of coherence on the system, we focus on two different approaches to the current calculation: the first restricts to the occupations $\n{i}{\kvv}$ to consider a rate equation approach - the so-called Wannier-Stark hopping model - and the second takes the coherences $\f{i}{j}{\kvv}$ into account.

We start with the rate equation approach, where in effect the electrons are counted as they cross an interface at a fixed point in the growth direction of the structure due to scattering between states which are localized at different spatial parts of the structure. The WSH current is just given by the application of Fermi's golden rule to each set of states. For the electron-phonon interaction, it is determined via ($n_{-\qv} = n_{\qv}$)
\ba \label{WSHcurr}
&&J_{\rm WSH}^{\pm}(t) = -2 \frac{2 \pi}{\hbar A L} \sum\limits_{i j} \sum\limits_{\kvv,\qv} |\g{i}{j}{\qv}|^2 \delpm{i}{j} \nonumber\\
&&\hspace*{0.7cm} \times \left( n_{\qv} + \frac{1}{2} \mp  \frac{1}{2} \right) (d_{i i} - d_{j j}) \n{i}{\kvv} (1 - \n{j}{\kvv+\qvv}),
\ea
where $J^{\pm}$ denotes the absorption/emission of a phonon, $d_{i i} = -e z_{ii}$ is the expectation value of $z$ for the WS state $|i\rangle$, and the sum $i,j$ is carried out over all states. The factor $2$ arises due to spin degeneracy and $L = N L_{\rm per}$ is the length of the structure with $N$ periods of length $L_{\rm per}$. The current is fully determined by the diagonal elements of the density matrix $\n{i}{\kvv}$, considering jumping of electrons from one state to the next, and reaches its steady state value as determined by the relaxation in Fig.~\ref{relax} when restricting to Boltzmann dynamics. An analogous expression is found for the elastic impurity scattering. While Eq.~(\ref{WSHcurr}) can be motivated by hopping of electrons between different positions $z_{ii}$ and $z_{jj}$, it is in fact an approximation for the full current driven by coherences, Eq.~(\ref{fullCurr}), see Ref.~\onlinecite{Wacker:PhysStatusSolidiC:08}.

When deriving the current microscopically from the current operator $\hat{J} = -e/(2 m_0) \fieldOpp [\hat{p} + e A(t)] \fieldOpm + \Hc$, one arrives at another picture of the current, where it is the coherence between states which carries the current flowing through the structure. Here, the current is given completely via the nondiagonal elements of the density matrix $\f{i}{j}{\kvv}$, which in growth direction yields
\be \label{fullCurr}
J_{\rm full}(t) = -2 \frac{e}{A L} \sum\limits_{i j} \sum\limits_\kvv v_{i j} \f{i}{j}{\kvv},
\ee
where $v_{i j} = \frac{1}{2} \langle i| [\hat{p}_z/m(z) + e A(t)/m(z)] + {\rm H.c.} |j\rangle$ are the velocity matrix elements and the factor $2$ again arises due to spin degeneracy. For the calculation, the polarizations $\f{i}{j}{\kvv}$ are initially set to zero. The finite occupations in the subbands lead to a stationary current which is fully determined by the scattering in the structure.\\

% \begin{figure}
% \begin{centering}
% \includegraphics[width=0.49\linewidth]{pics/currentKumarPhDot.eps}
% \includegraphics[width=0.49\linewidth]{pics/currentWSHfull.eps}
% \end{centering}
% \caption{(Color online) (left) Current evolution of the full model considering the different scattering mechanisms and (right) current evolution comparing the WSH and the full current model for a barrier width $b$ = 5.5 nm at $T$ = 10 K. {\bf *** redo both figures !!! ***}}
% \label{current}
% \end{figure}
% Fig.~\ref{current} shows the evolution of the currents, starting from the equally distributed subband populations as described in the previous section. In Fig.~\ref{current}(left), the evolution of the full current is shown for the different scattering mechanisms. Since the inelastic scattering with LO phonons restricts the interband scattering, a stationary current is reached on a much longer time scale than for the elastic impurity scattering which allows for a more efficient "thermalization" between the subbands. \commentsShort{is this true? what are the intra/inter scattering rates for ph,imp?} In Fig.~\ref{current}(right), the WSH and the full current evolutions are compared. While the full current is initially zero and dynamically obtains its steady state value, the WSH current is calculated directly from the subband occupations and thus takes a finite value for all times.

\begin{figure}
\includegraphics[width=0.9\linewidth]{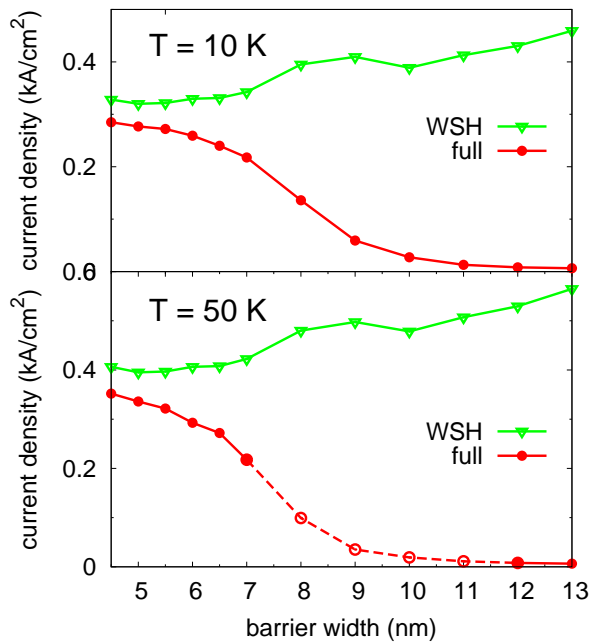}
\caption[]{\label{currBarr}(Color online) Stationary current for both the WSH and the full calculation at resonance for varying barrier widths at $T$ = 10 K, 50 K.\footnotemark[1]}
\end{figure}

We now carry out current calculations using both the WSH and the full approach for varying barrier widths. The result is shown in Fig.~\ref{currBarr}.
The WSH current remains approximately constant for varying barrier widths. This is due to the fact that at resonance, the injector and the upper laser level form a pair of binding/antibinding states whose nature hardly changes with $b$. Thus, the hopping current is not affected by the barrier width, see also Ref.~\onlinecite{Callebaut:JApplPhys:05}. The slight increase of the current may be attributed to the impact of the other levels which change slighly with decreasing bias, which is needed to keep the two tunnel-coupled states in resonance.

In contrast, the full calculation [Eq.~(\ref{fullCurr})] based on the coherences shows the expected drop of the current with barrier thickness. We first focus on the case $T$ = 10 K. For small barrier widths (where $2 \Omega \gtrsim \Gamma$), the two stationary current results are approximately the same and show a qualitatively similar behavior, i.e. the slope of the two curves are almost parallel. In this regime, the WSH and the full model both describe the stationary current well. The slightly smaller value for the full model stems from the fact that only the main injection barrier is investigated systematically here. Resonant tunneling at the other barriers in the structure is not considered, so that a similar argumentation as applied in the following can also be applied to them. Thus, it is expected that a constant offset can occur, probably due to the extraction barrier where the two delocalized states (lower laser level, extraction level) are also close to resonance.

For large barrier widths (where $2 \Omega \lesssim \Gamma$), the full current decreases until it almost vanishes in the limit of large barrier widths. This decrease is the physically expected behavior, as the growing injection barrier width restricts the current flowing across the barrier and thus through the whole structure. For larger barrier widths, the growing localization of the charge in the injection region leads to the fact that the Wannier-Stark states which are delocalized across the injection barrier no longer constitute a ``good'' basis to describe the states of the system. For this reason, the coherence between these two states becomes important in the stationary state, whereas it is negligible for smaller barrier widths. This can be seen in the relaxation dynamics as well, where for large barrier widths, the influence of the coherences on the populations, mainly of the injector and the upper laser state, becomes very important, and thus a pure Boltzmann relaxation description fails for the stationary populations.

For the case of $T$ = 50 K, the situation is different. For small barrier widths, the drop of the full current is more pronounced than for $T$ = 10 K, showing that even for those barrier widths the WSH model for the current fails. 
%Thus, for the entire parameter regime considered in the calculations, the WSH model for the current fails. 
We should note that this difference is not as pronounced as would be expected; however, since we are interested in qualitative results, it suffices to find that the onset of deviation between the WSH and the full calculation is shifted to lower barrier widths. It should be noted that for all barrier widths here, the relation $2 \Omega \ll \Gamma$ is fulfilled.

It is thus found that in the stationary current calculations, the inclusion of coherence becomes important if $2 \Omega \lesssim \Gamma$ under resonance. For $2 \Omega \gtrsim \Gamma$, the WSH approach to the current calculation is a good approximation, and thus the inclusion of coherence is negligible. For $2 \Omega \lesssim \Gamma$, only the full model shows the physically expected decrease of the current as the tunnel coupling decreases since the coherence between the injector and the upper laser state becomes important to describe the localization of the eigenstates. The decrease of the tunnel coupling across the main tunneling barrier {\it limits} the current flowing through the structure. This limitation is not reproduced by the WSH approach. It should be noted that this result has already been discussed before in Ref.~\onlinecite{Callebaut:JApplPhys:05} for the QCL and in Ref.~\onlinecite{Wacker:PhysRevLett:98} for superlattices.

\subsection{Ultrafast optical dynamics}

We now turn to the nonlinear optical response due to an ultrafast external perturbation of the laser structure from stationarity. Here, the situation is not as clear as in the transport case. Obviously, the light-matter interaction as such is a coherent process as the vector potential couples to the nondiagonal elements of the density matrix $\f{i}{j}{\kvv}$. However, it is more interesting to consider the dynamics after the passage of the pulse, where the signal is typically dominated by incoherent scattering.

As discussed in Sec.~\ref{secIIIB}, coherences between Wannier-Stark states are important in the regime of large barrier widths and high temperatures, where the stationary states are coherent superpositions of the WS states; they describe the tunneling through the barriers and thus the return to the stationary state. This is expected to remain important in the optical dynamics in order to describe the return to the steady state after optical excitation. However, the comparatively large scattering with respect to the tunneling (see Fig.~\ref{omegaGamma}) is expected to destroy all observable coherent effects in this regime.  The question arises whether coherent optical effects after ultrafast optical excitation can be observed in the opposite regime, where coherences were shown not to be important in the stationary transport. Recently, optical pump-probe measurements in mid-infrared\cite{Eickemeyer:PhysRevLett:02,Eickemeyer:PhysicaB:02,Woerner:JPhys:CondensMatter:04,Kuehn:ApplPhysLett:08} and THz QCLs\cite{Darmo::08} have shown an oscillatory behavior attributed to (coherent) resonant tunneling through the injection barrier. In this section, this effect is investigated with respect to the relation between the tunnel coupling and the level broadening.

We begin by considering the linear response of the system to characterize the laser. Then, we analyze the nonlinear optical response in form of the experimentally studied pump-probe signals to study the regime of importance and the role coherence plays in the optical signal. We complement this study by looking at the optically induced charge as well as the subband population dynamics.

\begin{figure}
%\includegraphics[width=0.49\linewidth]{pics/spectrumTemp.eps}
% \begin{minipage}{0.60\linewidth}
\includegraphics[width=0.8\linewidth]{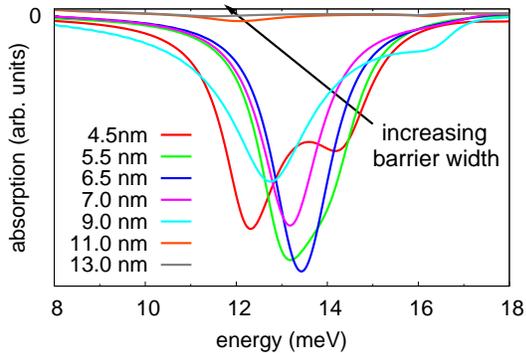}
% \end{minipage}
% \begin{minipage}{0.39\linewidth}
% \includegraphics[width=\linewidth]{pics/PPsetup.eps}
% \end{minipage}
%\caption{(Color online) (left) Temperature dependence and (right) injection barrier width dependence at $T$ = 10 K of the absorption spectrum. {\bf *** more T-data ! ***}}
\caption{\label{spectrum}Dependence of the absorption spectrum on the injection barrier width at $T$ = 10 K.% (right) Setup of the pump-probe response: A strong pump pulse saturates the gain transition, while the weak probe pulse tests the response at a fixed frequency and delay time.
}
\end{figure}
The linear absorption of the structure is calculated via the complex susceptibility 
\be
\chi(\omega) = \frac{1}{\eps_0} \frac{\delta J(\omega)}{\omega^2 A(\omega)}.
\ee
$\delta J(t) = J(t) - J_0$ denotes the change of the current density induced by the optical field with vector potential $A(\omega)$, where $J_0$ is the stationary value of the current density determined in the previous section and $J(t)$ is given by Eq.~(\ref{fullCurr}).
% \be
% J(t) = -2 \frac{e}{L} \left[ \frac{1}{A} \sum\limits_{i j} \sum\limits_\kvv v_{i j} \f{i}{j}{\kvv} - \frac{e}{m} A(t) n_{\rm tot} \right],
% \ee
% with the two-dimensional total electronic density per period $n_{\rm tot}$ which is just the doping density, $n_{\rm tot} = n_{\rm 2d}$. 
From this, the absorption $\alpha(\omega) \sim \omega {\rm Im} \chi(\omega)$ is calculated.
%The spectrum at different temperatures and for the injection barrier width as in the original QCL design is shown in Fig.~\ref{spectrum}(left). A strongly temperature-dependent gain is observed at $E \approx 14$ meV which decreases with temperature and almost vanishes at room temperature.

In Fig.~\ref{spectrum}, the absorption spectrum of the gain transition is shown for different injection barrier widths. Due to the anti-crossing at resonance, the spectrum shows a double-peak structure which can be resolved for small barrier widths where the tunnel splitting is sufficiently large. For larger barrier widths, the two peaks merge to form a single resonance for sufficiently small tunnel splitting energies compared to the dephasing of the transition. Due to this splitting, it is {\it a priori} unclear which of the transitions is to be considered the optical gain transition. In the following considerations, the system is excited resonantly on the transition yielding the larger peak gain which here is the lower gain transition frequency, and the accompanying optical dynamics is investigated. The second peak appearing for $b$ = 9.0 nm at $\Delta E \approx$ 16 meV arises from an enhanced dipole moment between the two resonant states $1',5$ and the extractor state $3$.

When increasing the barrier width further, the coherence between the injector and the upper laser state becomes important, so that the stationary state of the system is in a linear superposition of the two WS states, localized in the injector region. This coherence can lead to a strong decrease or even a vanishing of the gain. To illustrate this, it is helpful to consider a simple three-level system consisting of the two resonant states (injector $5$ and upper laser state $1'$) and the lower laser state $4$. We assume that the coherences between the lower laser state and other two states is small compared to the stationary tunneling coherence at the time of the probe pulse $t_0$, which is taken as a $\delta$-pulse, i.e. $|f_{4 1'}(t_0)|, |f_{4 5}(t_0)| \ll |f_{1' 5}(t_0)|$. Solving the semiconductor Bloch equations, the gain at the laser frequency is then given by (under the assumption $|E_5 - E_{1'}| \ll |E_5 - E_4|$)
\be
\alpha \sim -\left\{ d_{4 5}^2 (n_{5} - n_4) + d_{4 1'}^2 (n_{1'} - n_4) + 2 d_{4 5} d_{4 1'} \Real \left[ f_{1' 5}(t_0) \right] \right\}
\ee
with the constant level densities $n_i$. If the system is now in the state $|\Psi\rangle \approx |5\rangle + |1'\rangle$ ($d_{4 5} \approx -d_{4 1'}$) which is localized in the injection region, as is the case for large barrier widths, we have $n_5 \approx n_{1'}$. The gain is then given by the stationary inversion of the states $n_{5} - n_4$ as well as a further term $-2 \Real [f_{1' 5}(t_0)]$ containing the coherence between the two resonant states,
\be
\alpha \sim -d_{4 5}^2 \left\{ 2 (n_{5} - n_4) - 2 \Real \left[ f_{1' 5}(t_0) \right] \right\}.
\ee
Typically, the first term containing the inversion determines the optical spectrum. In the presence of a strong stationary coherence $f_{1' 5}(t_0)$, the gain can be strongly reduced or even vanish, even in the presence of a strong inversion, which is the case for all barrier widths considered. The stationary state as a superposition of the two WS states is fully localized in the injector, leading to a very small dipole overlap with the lower laser state which is witnessed in the vanishing gain.\\

\begin{figure}
\includegraphics[width=\linewidth]{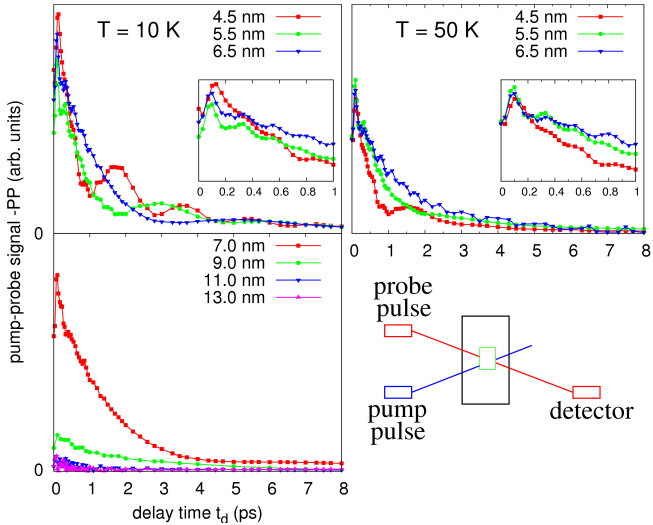}
\caption{\label{PP}(Color online) Pump-probe signals after ultrafast strong pumping and subsequent linear probing at the laser energy for different delay times $t_d$ and barrier widths at $T$ = 10 K, 50 K. Both pump and probe pulses are chosen to be resonant on the laser transition. Insets: Pump-probe signals for small delay times.}
\end{figure}

% \begin{figure}
% \begin{centering}
% \includegraphics[width=0.6\linewidth]{pics/PPsetup.eps}
% \end{centering}
% \caption{(Color online) Setup of the pump-probe response: A strong pump pulse saturates the gain transition, while the weak probe pulse tests the response at a fixed frequency and delay time.}
% \label{PPsetup}
% \end{figure}
For the nonlinear optical dynamics, we focus on the pump-probe signals which have been experimentally measured recently.\cite{Eickemeyer:PhysRevLett:02,Eickemeyer:PhysicaB:02,Woerner:JPhys:CondensMatter:04,Darmo::08,Choi:Phys:RevLett:08,Choi:ApplPhysLett:08,Kuehn:ApplPhysLett:08} An ultrafast nonlinear pump pulse resonant on the laser transition excites the sample, leading to a gain saturation at the laser energy, and a subsequent weak probe pulse tests the laser transition as a function of the delay time between the two pulses (see Fig.~\ref{PP}). Corresponding to the experiments, a 170 fs 1$\pi$ pump pulse (P) and a (in comparison to all optical transitions of interest) spectrally broad probe pulse (Pr, with $T$ = 50 fs) are used. In contrast to a standard differential transmission spectrum where the complete energy range is recorded,\cite{Haug::04} the focus is on a monochromatic measurement, yielding the absorption at a fixed energy for different delay times. The experimental separation of pump and probe pulses which is done via directional filtering of the response is carried out here in such a way that the response to only the pump pulse $\delta J_{\rm P}$ is subtracted from the response to both the pump and probe pulses $\delta P_{\rm P+Pr}$, yielding the purely linear response after the influence of the pump pulse whose direct response is filtered out. Thus, we consider for the absorption
\be \label{pumpProbeAlpha}
\alpha(t_d,\bar{\omega}) \sim \Imag \left[ \frac{\delta J_{\rm P+Pr}(t_d,\bar{\omega})-\delta J_{\rm P}(\bar{\omega})}{\bar{\omega} A_{\rm Pr}(t_d,\bar{\omega})} \right],
\ee
with the delay time $t_d$, the fixed pump and probe energy $\bar{\omega}$ (which is equal to the laser energy here), and the probe field $A_{\rm Pr}$. $\delta J_i(\bar \omega)$ is the Fourier transform of the corresponding simulated time signal $\delta J_i(t)$ taken at the fixed frequency $\bar{\omega}$. In Ref.~\onlinecite{Richter:PhotosynthRes:08}, it is shown that the nonlinear absorption according to Eq.~(\ref{pumpProbeAlpha}) describes the signal contribution measured in the probe direction. While the signal contains all coherent effects also in the limit of delay times shorter than or comparable with the dephasing times, additional coherent effects are included for short delay times which are not measured experimentally due to directional filtering; these effects are expected to be small. Our result reproduces well known studies in the probe direction, for instance Ref.~\onlinecite{Koch:JPhysC:88}. The differential pump-probe response is then given by $\exp \{-[\alpha(t_d) - \alpha_0(t_d)] L \} - 1 \sim -\alpha(t_d) + \alpha_0(t_d) \equiv {\rm PP}(t_d)$, where $\alpha_0(t_d)$ denotes the linear response without the pump pulse and it is assumed that $\alpha L \ll 1$ where $L$ is the length of the structure. For reasons of presentation, we plot the negative differential pump-probe response. Then, positive signals correspond to a decreased gain compared to the stationary reference gain value.
% As expected, our calculations satisfy ${\rm PP}(t_d) \rightarrow 0$ for $t_d \rightarrow -\infty$.

\begin{figure}
\includegraphics[width=0.49\linewidth]{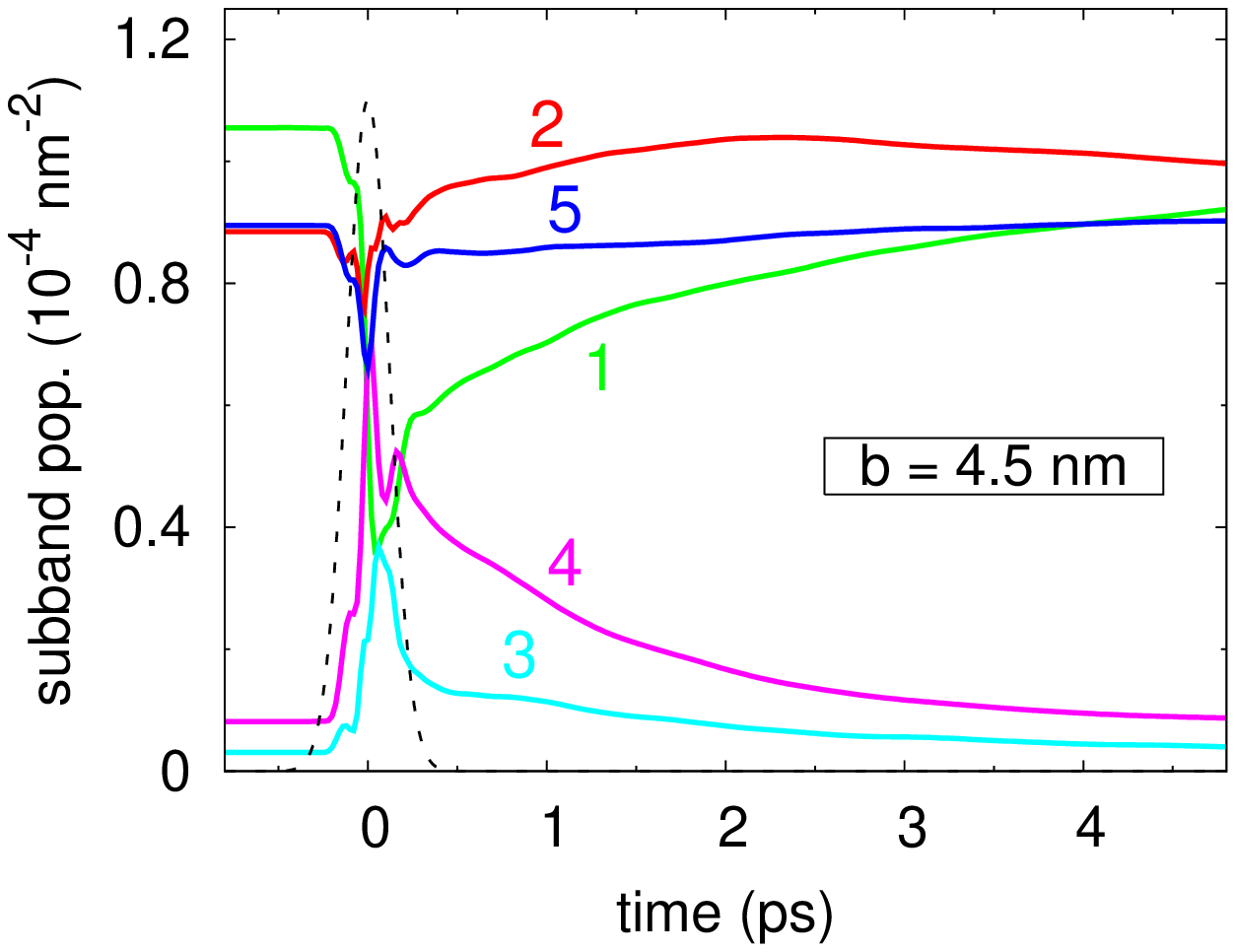}
\includegraphics[width=0.49\linewidth]{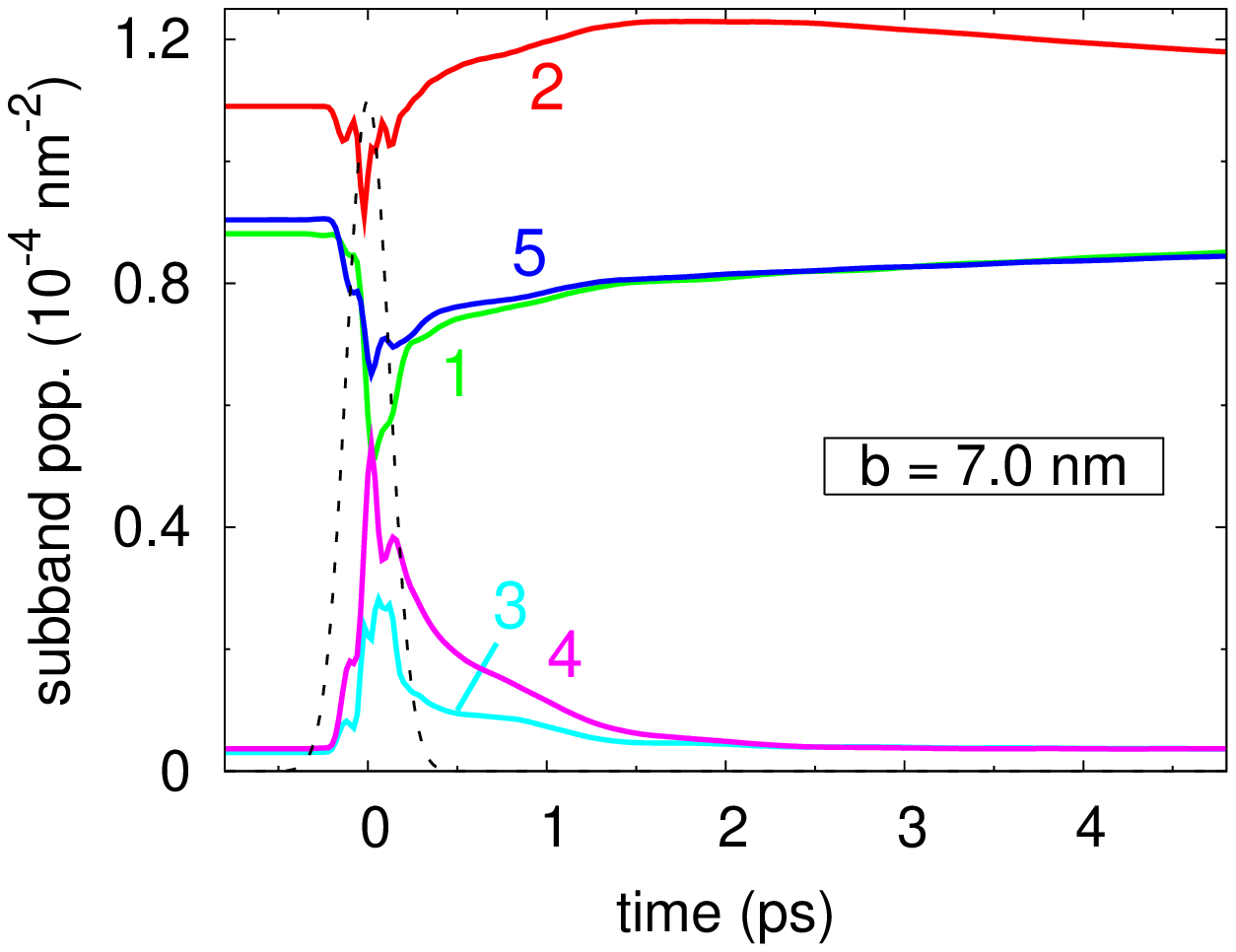}
\caption{\label{popDyn}(Color online) Optically induced population dynamics $\n{i}{}(t)$ for the barrier widths $b$ = 4.5, 7.0 nm at $T$ = 10 K, showing coherent Rabi oscillations as well as subsequent scattering-induced relaxation back into the steady state. The pump pulse parameters are the same as for the pump-probe results (the pump pulse is shown by the dashed line in the figures).}
\end{figure}

The pump-probe response for different barrier widths and at $T$ = 10 K, 50 K is shown in Fig.~\ref{PP} (see also Fig.~\ref{popDyn} for the corresponding population dynamics caused by the pump pulse for $b$ = 4.5, 7.0 nm at $T$ = 10 K).
The ultrafast pump pulse saturates the gain transition, leading to a strong absorption at this energy (positive pump-probe signal). The following return to the steady state shows two main features: (i) A decay of the signal on a time scale of a few picoseconds. This feature is due to the incoherent scattering which leads to a return to the steady state, yielding the original gain for the laser transition. (ii) For smaller barrier widths, an oscillatory modulation of the signal whose amplitude decreases for growing barrier widths, while its period of the order of picoseconds increases correspondingly. In addition, fast oscillations with a period on the order of 0.1 ps are visible which are addressed briefly below, but which are not the focus of the article.

The oscillatory feature is a coherent effect in the dynamics after the passage of the pulse: The pump pulse depopulates the upper laser state locally in the excited region. At resonance condition, such a locality is represented by a coherent superposition of the levels $1'$ and $5$. This causes coherent charge oscillations between the two superpositions which are localized in the injector and in the active region,\cite{Kazarinov:SovPhys--Semicond:72} leading to a modulation of the coherence between the two laser states and thus to an oscillation of the gain signal. The oscillation hardly causes any changes in the WS subband populations and no oscillating inversion as will be discussed later in this section. Thus, the oscillation is a coherent effect. The charge oscillations are visible for the case that the tunneling period $T_{\rm osc} \sim 1/\Delta E$ is less than the scattering period $T_{\rm scatt} \sim 1/\Gamma$, i.e. $T_{\rm osc} \lesssim T_{\rm scatt}$. Let us first focus on $T$ = 10 K. Here, gain oscillations are strongly visible for $b$ = 4.5 nm, 5.5 nm where $2 \Omega \gtrsim \Gamma$, i.e. $T_{\rm osc} \lesssim T_{\rm scatt}$. Already for $b$ = 6.5 nm, the oscillations become hardly visible and fully vanish for $d \geq$ 7.0 nm, where $T_{\rm osc} \gg T_{\rm scatt}$. For very large barrier widths, it is not possible to invert the gain transition due to the strong coherence between the injector and the upper laser state. The two states are now in a superposition state localized in the injector,  and thus a dipole interaction via the electric field between the lower laser state and the upper state is not possible since the charge is localized outside the active region. Thus, no charge inversion occurs (compare the discussion of the gain for large barrier widths in Fig.~\ref{spectrum}). At $T$ = 50 K, $2 \Omega \lesssim \Gamma$, i.e. $T_{\rm osc} \gtrsim T_{\rm scatt}$ is fulfilled for all barrier widths (cf. Fig.~\ref{omegaGamma}), and thus only very weak gain oscillations are found for $b$ = 4.5 nm. Again, for large barrier widths, no inversion occurs, leading to an almost vanishing pump-probe signal. It should be noted that the fast oscillations on a time scale of around 300 fs seen in the insets of the pump-probe signals for small delay times $t_d \lesssim$ 1 ps can be attributed to tunneling due to the coherence between the lower laser state and the two resonant states which show an energy splitting of $\Delta E \approx$ 13 meV, corresponding to an oscillation period of $T_{\rm osc} \approx$ 300 fs. Due to the changing applied bias and the correspondingly changing laser energy, the oscillation period changes slightly between $T_{\rm osc} \approx$ 200--300 fs.

Experimentally, gain oscillations have been observed in a mid-infrared laser in Refs.~\onlinecite{Eickemeyer:PhysRevLett:02,Eickemeyer:PhysicaB:02,Woerner:JPhys:CondensMatter:04,Kuehn:ApplPhysLett:08} and in a THz laser in Ref.~\onlinecite{Darmo::08}. In the case of the mid-infrared laser, pronounced gain oscillations were found up to relatively high temperatures, including a gain overshoot when probing close to the resonance. This effect depends strongly on the strength of the scattering mechanisms, specifically on the depletion of the lower laser subband and the strong coupling through the injection barrier, and has not been observed in our calculations. According In Ref.~\onlinecite{Woerner:JPhys:CondensMatter:04}, the lifetime of the superposition of the two resonant states $T_{\rm scatt} \approx$ 1 ps which is much longer than the observed oscillation of $T_{\rm osc} \approx$ 500 fs, and thus $2 \Omega > \Gamma$ which is the regime where gain oscillations are expected. In Ref.~\onlinecite{Choi:ApplPhysLett:08}, time-resolved pump-probe differential transmission measurements did not show signs of gain oscillations. As the authors argue in the paper, this is due to the very short scattering rates due to Coulomb scattering ($T_{\rm scatt} <$ 100 fs) which is much smaller than the tunneling oscillation period ($T_{\rm osc}$ = 517 fs), and thus no gain oscillations are expected.

\begin{figure}
\includegraphics[width=0.49\linewidth]{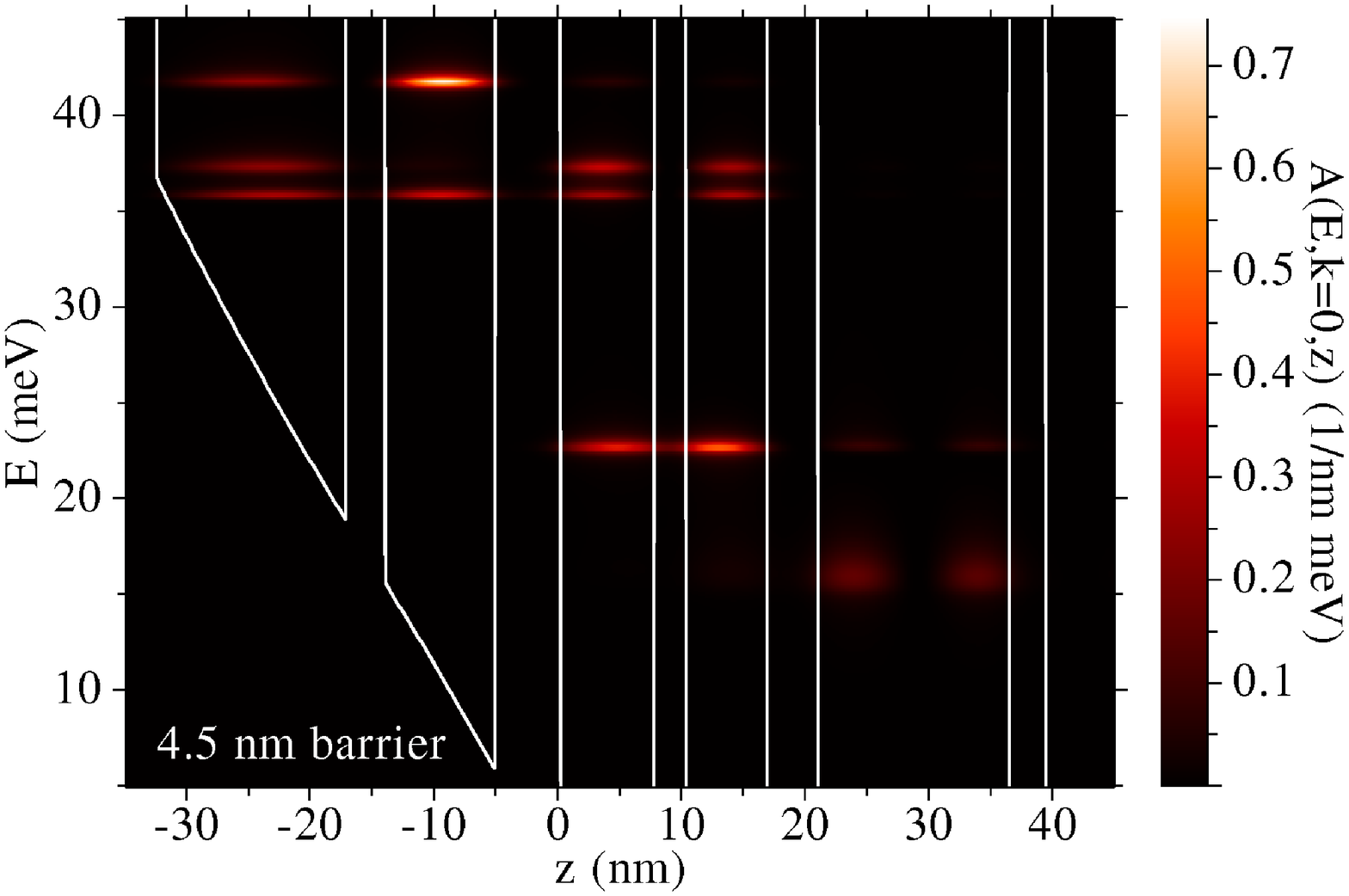}
\includegraphics[width=0.49\linewidth]{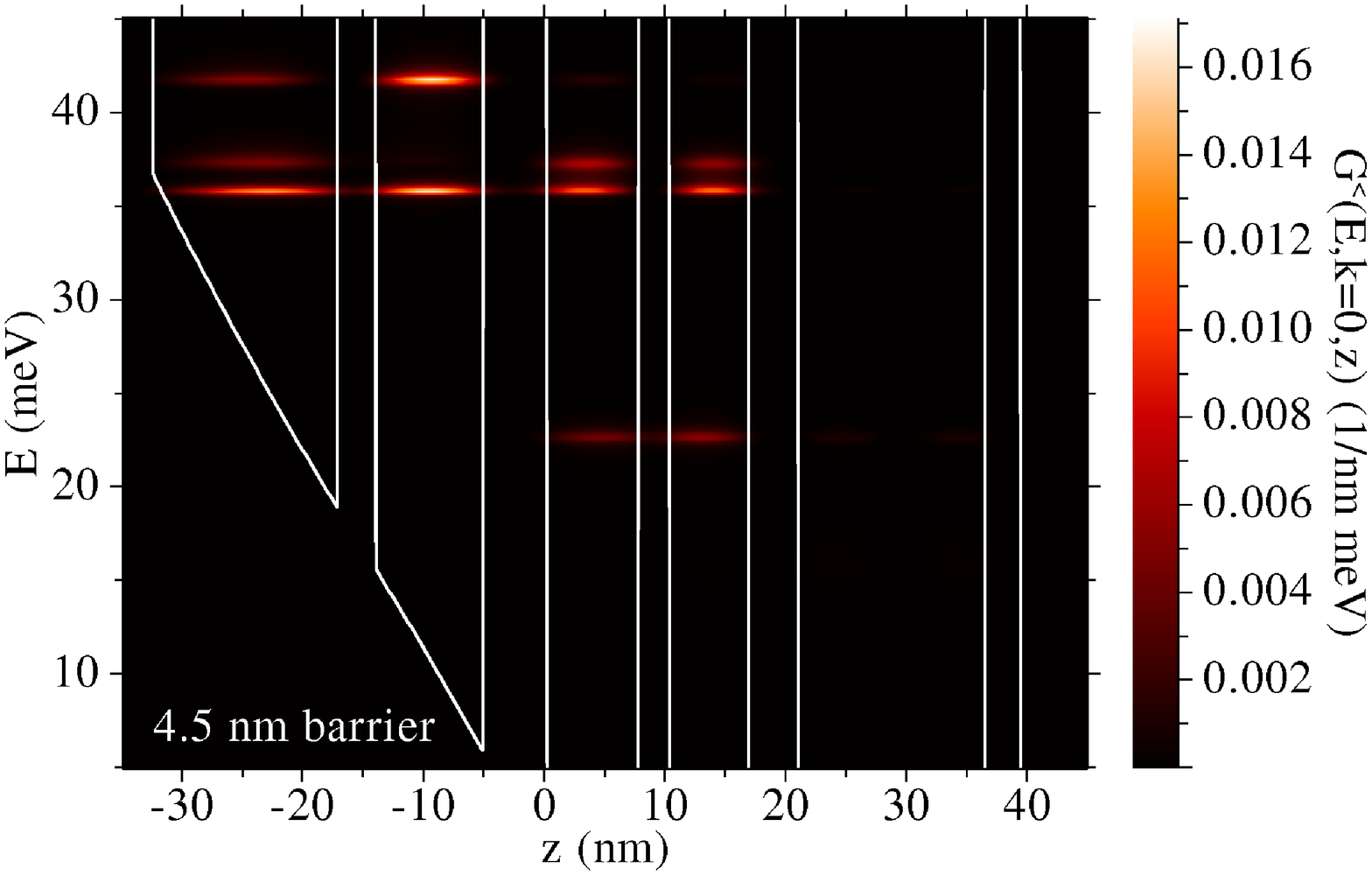}\\
\includegraphics[width=0.49\linewidth]{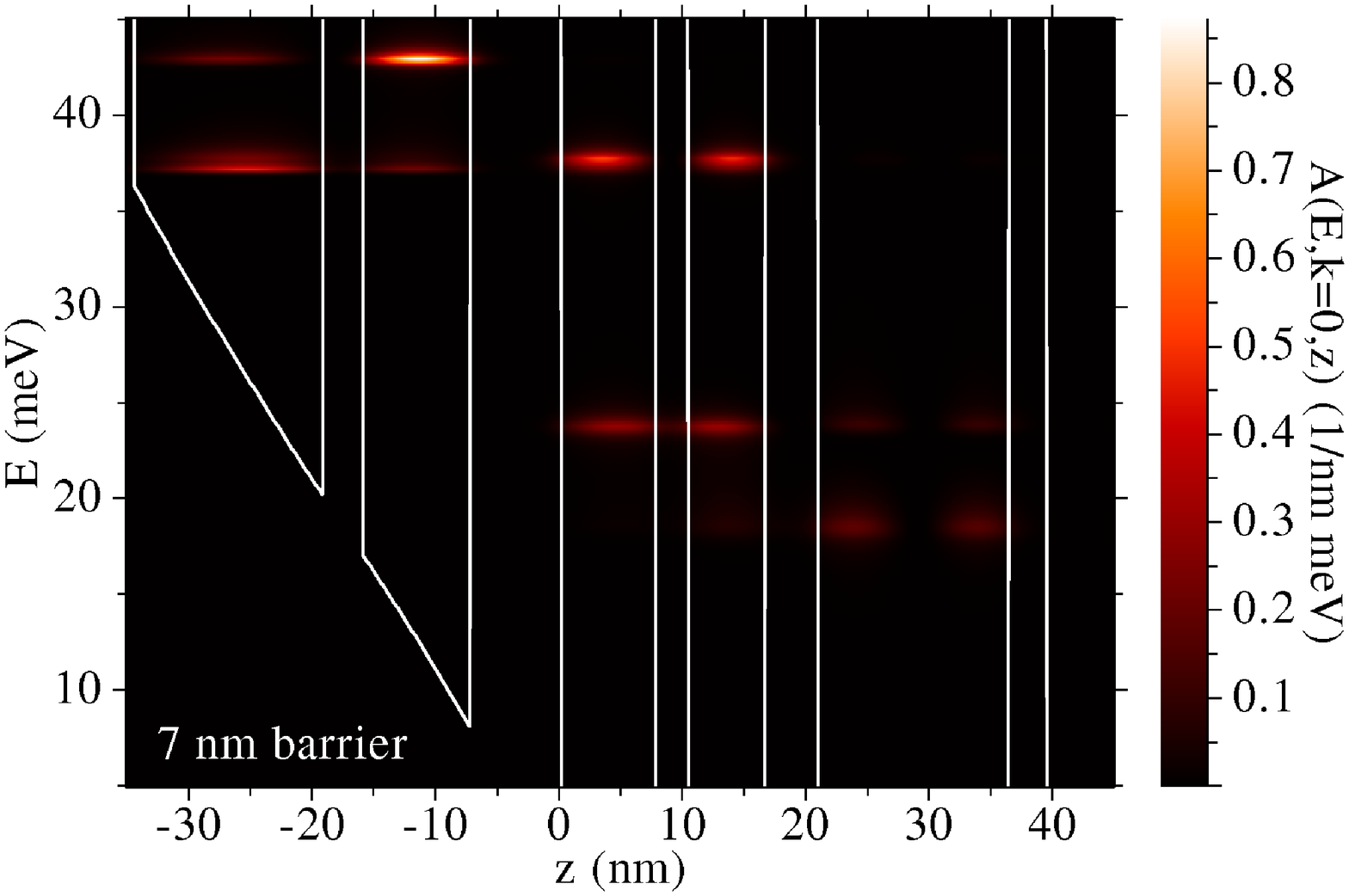}
\includegraphics[width=0.49\linewidth]{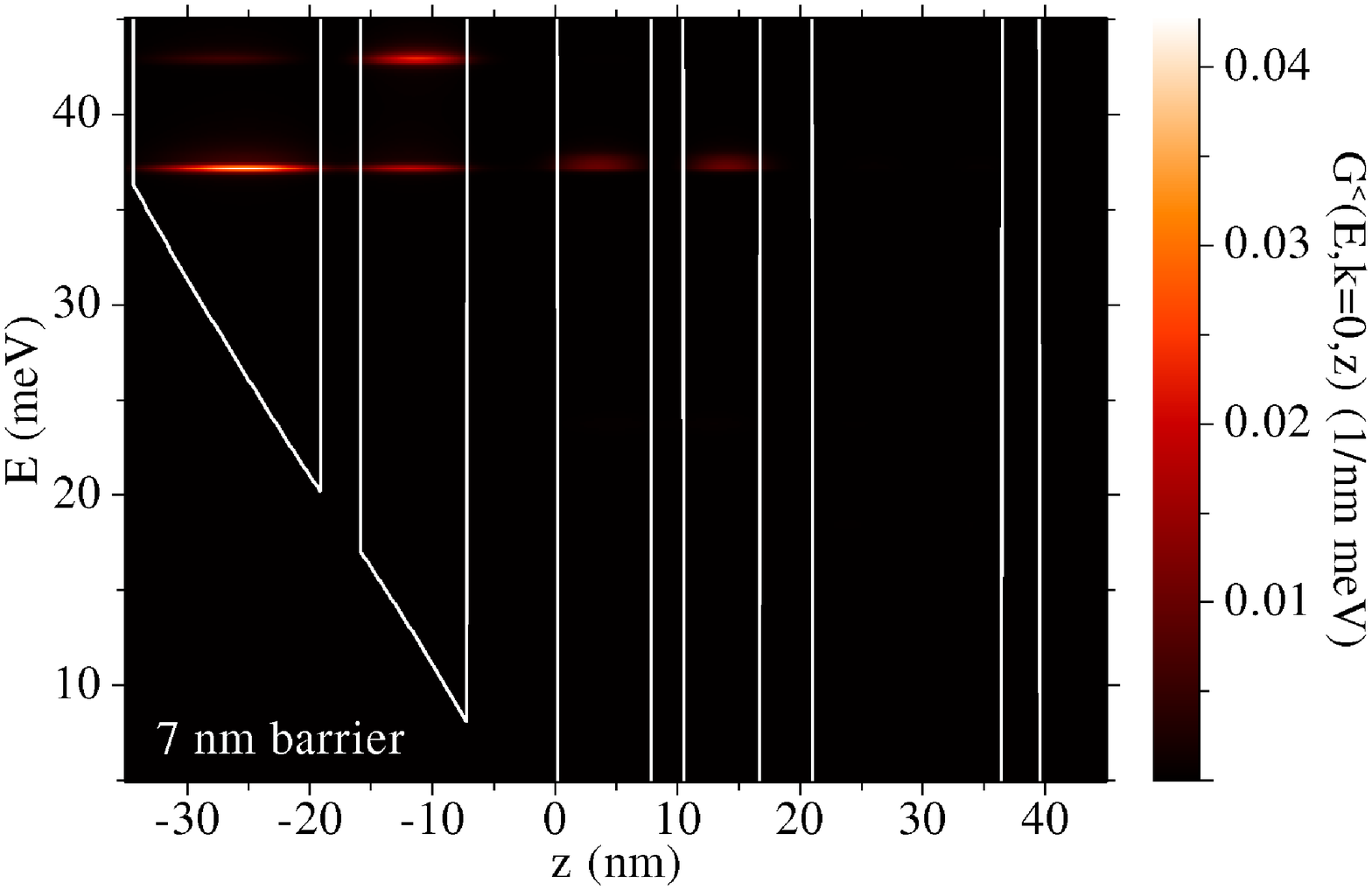}
\caption{\label{DOS}(Color online) (left) Spectral function and (right) the imaginary part of the lesser Green's function for two different barrier widths at $T$ = 10 K.}
\end{figure}
The interpretation of the oscillations in the pump-probe signals being caused by coherent charge transfer across the injection barrier is strengthened by considering the spatio-energetic structure of the stationary state obtained from our Green's function model.\cite{Lee:PhysRevB:06} In Fig.~\ref{DOS}(left), the spectral function $A(E,k=0,z)$ which describes the density of states for vanishing in-plane momentum is shown. The peaks roughly correspond to the WS wave functions $|\xi_i(z)|^2$ with an energetic width $\Gamma_i$. For $b$ = 4.5 nm, the binding and antibinding combination of the injector and the upper laser level can be resolved around $E \approx$ 36 meV, while this is not the case for $b$ = 7.0 nm as $\Delta E < \Gamma$. The right figures show the lesser Green's function $\Imag [ G^<(E,k=0,z) ]$ describing the corresponding carrier density. For $b$ = 4.5 nm, the carrier density qualitatively follows the spectral function corresponding to two occupied levels in the stationary state. Thus, when perturbing the stationarity on an ultrashort time scale, a superposition of the two states leads to an oscillatory modulation of the gain. For $b$ = 7.0 nm, the carrier density does not follow the density of states; only the superposition of the two states, localized in the injector, is occupied at stationarity, and thus no superposition can be excited and no oscillations occur. The stationary distribution is already a superposition of the two WS states, which is also the reason why the inclusion of coherence is important to correctly describe the stationary transport of the system (see Fig.~\ref{currBarr}).

In addition to the spatio{\it energetic} resolution of the stationary charge density, it is insightful to consider its spatio{\it temporal} evolution during and especially after the excitation with an ultrafast laser pulse. To do this, we consider the spatio-temporally resolved electron density which is given by
\be
n(z,t) = 2 \frac{1}{A} \sum\limits_{i j} \xi_i^*(z) \, \xi_j(z) \sum\limits_{\kvv} \f{i}{j}{\kvv}(t),\label{elecDens}
\ee
where the time dependence is included in the density matrix $\f{i}{j}{\kvv}(t)$. In order to simplify the representation and to focus on the {\it optically induced} dynamics, the difference between $n(z,t)$ and the stationary electron density $n_0(z)$, $\Delta n(z,t) = n(z,t) - n_0(z)$, is considered in the following.
%It should be noted that in contrast to the {\it optically induced} perturbation of the carrrier dynamics considered here, the spatio-energetically resolved {\it stationary} current density was discussed recently in Ref. \onlinecite{Lee:PhysRevB:06}.
Similar calculations have been presented in Ref.~\onlinecite{Iotti:PhysRevLett:01}, where however no detailed modeling of the pump pulse was done.

\begin{figure}
\includegraphics[width=\linewidth]{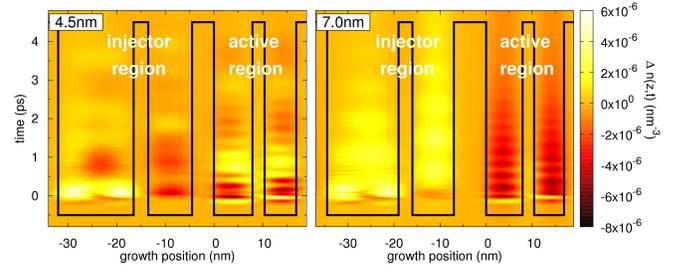}
\caption{\label{elDens}(Color online) Spatio-temporal evolution of the electron density of the QCL after ultrafast nonlinear optical excitation at $T$ = 10 K. The pump pulse parameters are the same as for the pump-probe results.}
\end{figure}

In Fig.~\ref{elDens}, the electron density evolution is shown for the same parameters of the pump pulse as used for the calculation of the pump-probe signal discussed above.
Shortly after the pulse has passed at $t \approx$ 0.3 ps, the density is depleted in the active region due to the fast nonradiative extraction from the lower laser level to the extractor state and into the injector.

After this, two qualitatively different behaviors are seen for the two barrier widths. For $b$ = 4.5 nm, the resonant tunneling across the injection barrier leads to a fast refilling of the upper laser state and thus to an increased density in the active region at $t \approx$ 1 ps. At $t \approx$ 2 ps, the density in the active region again decreases. Thus, the oscillation of charge which occurs on a time scale of 1.8 ps as detected by the probe pulse, see Fig.~\ref{PP}(left), can be directly seen in the electron density evolution. When considering the electron density without the polarizations, i.e. restricting to the diagonal elements of the density matrix for the calculation, or even just without including the coherences between different periods (not shown), this oscillation is not observed at all.

For the case of $b$ = 7.0 nm, a roughly monotonous return of the density depletion to the steady state is found, and thus no gain oscillations are expected, as verified in the pump-probe signals. Due to the weaker tunnel coupling between the injector and the active region, the charge transfer is slow so that the oscillation period induced by the tunnel coupling $T_{\rm osc}$ is larger than the scattering period $T_{\rm scatt}$, and thus charge oscillations cannot be observed. Additionally, there are fast oscillations in the active region on a time scale of around 300 fs which we already addressed in the discussion of the pump-probe signals. The oscillations in the {\it injector region} can be explained by coherences between the injector states where $E_2 - E_5 \approx$ 5 meV $\hat{=}$  850 fs.

It has already been mentioned that the charge oscillations are not seen in the electron density evolution when only the diagonal elements of the density matrix are used in Eq.~(\ref{elecDens}). Thus they constitute a coherent effect. This is better exemplified by considering directly the subband population evolution caused by the strong pump pulse excitation. We thus consider the dynamics of the WS populations $n_i(t) = 2/A \sum_\kvv \n{i}{\kvv}(t)$ under ultrafast optical excitation. It should be stressed here that the dynamical calculations are carried out with the full equation structure including both the diagonal and nondiagonal density-matrix elements. 
%The different coherence regimes in the subband population dynamics, from the coherent to the quasi-stationary response, and the importance of including coherence between states of different periods during the modeling of these processes has been studied in Ref. \onlinecite{Weber:ApplPhysLett:06}.
Again, the dynamics is considered for the same nonlinear excitation used in the pump-probe calculations. The results are shown in Fig.~\ref{popDyn}.

During the excitation with the pump pulse, the populations of the laser levels undergo Rabi flopping, while the other levels follow adiabatically.\cite{Binder:PhysRevLett:90,Cundiff:PhysRevLett:94,Hughes:PhysRevLett:98,Weber:ApplPhysLett:06} After the excitation, the populations return to the steady state due to the (incoherent) scattering processes. While for $b$ = 4.5 nm, the Rabi flopping is restricted mainly to the two resonantly excited subbands ($1'$,$4$), the subband flopping for $b$ = 7.0 nm includes both the two laser states ($4$,$1'$) and the injector level ($5$) since both transitions from the lower laser state are energetically very close due to the very small tunnel splitting. This weakens the oscillator strength of each transition, reducing the effective pulse area compared to the full population inversion seen for a 1$\pi$ pump pulse in an ideal two-level system. Still, in both cases, the gain transition is saturated, leading to absorption at this energy. An oscillating inversion of the upper und lower laser states corresponding to the oscillations in the pump-probe signals (Fig.~\ref{PP}) or in the spatio-temporal electron density evolution on a time scale of a few picoseconds (Fig.~\ref{elDens}) is found neither for $b$ = 4.5 nm nor for $b$ = 7.0 nm. As a matter of fact, if the dynamics are considered without scattering (not shown), no further population dynamics are found after the passage of the pulse, while the corresponding pump-probe signals show an undamped oscillation in the case of $b = 4.5$ nm. This illustrates again that the oscillations in the gain recovery are a coherent effect.

In summary, it is found that gain oscillations are observed where the tunnel oscillation period is sufficiently less than the scattering period, where the relation $2 \Omega \gtrsim \Gamma$ is valid. The tunnel coupling leads to pronounced  coherent charge transfer between the injector and the active region as witnessed in the pump-probe signals and the spatio-temporally revolved electron density evolution. Considerations of quantities which do not directly relate to the coherence, such as the dynamics of the WS populations, show that the oscillations observed in the pump-probe signals constitute a coherent effect which requires a fully coherent theory to be observed. It should be noted here that compared to the current calculations, where a {\it stationary localization} of charge leads to the drop of the current signal, it is here the {\it dynamical transfer} of charge between different locations in the laser structure which constitutes the coherent effect. Thus, the inclusion of coherence becomes essential in the nonlinear optical dynamics for the regime where $2 \Omega \gtrsim \Gamma$ to describe the observed coherent effects. Here, the coherence in the system {\it drives} the gain oscillations, while for $2 \Omega \lesssim \Gamma$, the scattering destroys the coherent effects.

% In order to investigate the effective charge transfer between the two regions, we have performed a spatial averaging over both regions which is shown in Fig.~\ref{WP}. It can be clearly seen that charge is transferred between the injector and the active region. Without a consideration of wave packets spatially localized in one specific period which would require an spatially limited excitation of the QCL, it is however not possible to find out whether this is ballistic charge transfer through the structure or whether there is charge oscillating from the injector into the active region and back into the injector of the prior period (coherent charge transport).

\section{Conclusion}

We have established a microscopic quantum theory of scattering for quantum cascade lasers to treat stationary transport as well as nonlinear optical dynamics. A key issue is the importance of coherence, i.e. the nondiagonal elements of the density matrix, and the role which it plays in different parameter regimes of the tunnel coupling $\Omega$ between the two states across the main injection barrier and the mean level broadening $\Gamma$ of the states. In the transport regime, we have compared a rate equation (WSH) approach to the current with our microscopic theory at resonance condition and found that for $2 \Omega \gtrsim \Gamma$, the WSH approach is a good approximation. For $2 \Omega \lesssim \Gamma$, only the full approach shows the physically expected decrease of the current for decreasing tunneling across the main injection barrier, whereby the included coherence {\it limits} the current flowing through the structure via a stationary localization of charge in the injection region. On the other hand, the consideration of ultrafast nonlinear optical pump-probe signals has shown that the inclusion of coherence is important to describe the experimentally observed coherent effects for $2 \Omega \gtrsim \Gamma$, where it {\it drives} the dynamical charge transfer between the injector and the active region, resulting in oscillations in the gain recovery. For $2 \Omega \lesssim \Gamma$, only the typical decrease of the pump-probe signal due to incoherent scattering is observed. It should be noted again that, in general, coherences are necessary to describe the return to the stationary state after optical excitation due to the strong interplay between transport and optics in this system. The spatially resolved electron density evolution in time shows the density oscillating between the injector and the active region for $2 \Omega \gtrsim \Gamma$. The consideration of the population dynamics of the levels do not show these oscillations, revealing that they are an inherent coherent effect resulting from the nondiagonal elements of the density matrix.

Thus, the inclusion of coherence is important in opposite parameter regimes in stationary transport and nonlinear optical dynamics: in the former, it becomes important for $2 \Omega \lesssim \Gamma$, while in the latter, it allows the observability of coherent effects for $2 \Omega \gtrsim \Gamma$. In this sense, the coherence acts in complementary ways: in the transport regime, it leads to a {\it limitation} of the current which is not reproduced by the rate equation approach. In the optical regime, the coherence {\it drives} the oscillatory modulation of the gain recovery which is not destroyed by the scattering. Thus, it is necessary to consider a fully coherent theory in order to describe the combined system of optics and transport in this nonequilibrium structure in the entire parameter range.

\begin{acknowledgments}
We acknowledge fruitful discussions with Rikard Nelander, Marten Richter, Stefan Butscher, Michael Woerner, and Klaus Reimann. This work was supported by the Deutsche Forschungsgemeinschaft (DFG) and the Swedish research council (VR).
\end{acknowledgments}

\appendix*
\section{Scattering equations}

\subsection{Electron-LO phonon interaction}

% In the following, the equations as well as the scattering rates for the electron-LO phonon interaction are presented.
The general form of the scattering equations in Born-Markov approximation is given by (for $\Bosee{\qv} = \Bosee{-\qv}$)
\begin{widetext}
\ba \label{elPhEqs}
\df{i}{j}{\kvv} &=&\frac{\pi}{\hbar} \sum\limits_{l m n} \sum\limits_\qv \Nph{l}{i}{m}{n}\\
&&\;\;\; \times \left\{ (\delta_{nj} - \f{n}{j}{\kvv}) \f{l}{m}{\kvv+\qvv} \left[ (\Bosee{\qv} + 1) \delp{n}{m} + \Bosee{\qv} \delm{n}{m} \right] \right. \nonumber\\
&&\left.\;\;\;\;\;\;\;\; - (\delta_{lm} - \f{l}{m}{\kvv+\qvv}) \f{n}{j}{\kvv} \left[ (\Bosee{\qv} + 1) \delm{n}{m} + \Bosee{\qv} \delp{n}{m} \right] \right\}\nonumber\\
&&+ \frac{\pi}{\hbar} \sum\limits_{l m n} \sum\limits_\qv \Nphconj{l}{j}{m}{n} \nonumber\\
&&\;\;\; \times \left\{ (\delta_{in} - \f{i}{n}{\kvv}) \f{m}{l}{\kvv+\qvv} \left[ (\Bosee{\qv} + 1) \delp{n}{m} + \Bosee{\qv} \delm{n}{m} \right] \right. \nonumber\\
&&\left.\;\;\;\;\;\;\;\; - (\delta_{ml} - \f{m}{l}{\kvv+\qvv}) \f{i}{n}{\kvv} \left[ (\Bosee{\qv} + 1) \delm{n}{m} + \Bosee{\qv} \delp{n}{m} \right] \right\}\nonumber,
\ea
\end{widetext}
where $\Nph{i}{j}{m}{n} = \g{i}{j}{\qv} \gconj{m}{n}{\qv}$. In the calculations, the considerations are restricted to terms linear in the density matrix since in the systems considered here, $|\f{i}{j}{\kvv}| \ll 1$ so that $|\f{i}{j}{\kvv} \f{l}{m}{\kvv}| \ll |\f{i}{j}{\kvv}|$.

The Boltzmann relaxation is determined via
\be
\dot{n}_{i \kvv}^{\rm (0)} = -\Gamma^{\rm out}_{i \kvv} n_{i \kvv} + \Gamma^{\rm in}_{i \kvv} (1 - n_{i \kvv}).
\ee
with the semiclassical Boltzmann in- and out-scattering rates given by
%%%% El-Ph Dichten Boltzmann %%%%
\ba
&\Gamma^{\rm in}_{i \kvv} &= \frac{2 \pi}{\hbar} \sum\limits_{\qv,l} |\g{i}{l}{\qv}|^2 \n{l}{\kvv+\qvv} \left[ (n_\qv + 1) \delp{i}{l} \right. \nonumber\\
&&\left. \hspace*{2.5cm} + n_\qv \delm{i}{l} \right], \\
&\Gamma^{\rm out}_{i \kvv} &= \frac{2 \pi}{\hbar} \sum\limits_{\qv,l} |\g{i}{l}{\qv}|^2 (1-\n{l}{\kvv+\qvv})  \left[ n_\qv \delp{i}{l} \right. \nonumber\\
&&\left. \hspace*{2.0cm} + (n_\qv + 1) \delm{i}{l} \right].
\ea
\subsection{Ionized doping centers - Impurity scattering}

% 
% In this section, the equations and the scattering rates for the interaction of the electrons with the \changeCW{ionized doping centers} are presented.
The general form of the scattering equations is given by
\begin{widetext}
\ba \label{elDotEqs}
&\df{i}{j}{\kvv} = -\frac{\pi}{\hbar} \sum\limits_{m n} \sum\limits_\qvv &\!\!\!\left[ \Mimp{m}{i}{n}{m} \f{n}{j}{\kvv} \delta(\eps_{n \kvv} - \eps_{m \kvv+\qvv}) - \Mimp{m}{i}{j}{n} \f{m}{n}{\kvv+\qvv} \delta(\eps_{j \kvv} - \eps_{n \kvv+\qvv})\right.\nonumber\\
&&\left.+ \Mimp{m}{j}{n}{m} \f{i}{n}{\kvv} \delta(\eps_{n \kvv} - \eps_{m \kvv+\qvv}) - \Mimp{m}{j}{i}{n} \f{n}{m}{\kvv+\qvv} \delta(\eps_{i \kvv} - \eps_{n \kvv+\qvv}) \right],
\ea
\end{widetext}
where the average $\Mimp{i}{j}{m}{n} = \langle \sum_{\bar{l},\bar{l}'} \V{\bar l}{i}{j}{\qvv} \V{\bar l'}{m}{n}{-\qvv} \rangle^c_S$ over the statistically distributed doping atoms is introduced. Since the semiclassical interaction is linear in the density matrix, all terms are taken along for this interaction.

% Again, we first focus on the occupations and then discuss the polarization dynamics.
% 
% \paragraph*{\changeCW{Occupation} dynamics}
% 
The semiclassical Boltzmann
% \be
% \dn{i}{\kvv} = -\Gamma^{\rm out}_{i \kvv} \n{i}{\kvv} + \Gamma^{\rm in}_{i \kvv} (1 - \n{i}{\kvv})
% \ee
in- and out-scattering rates are given by
\ba
&\Gamma^{\rm in}_{i \kvv} &= \frac{2 \pi}{\hbar} \sum\limits_{\qvv,l} M^{i l}_{i l} \deltoo{i}{l}  \n{l}{\kvv+\qvv},\\
&\Gamma^{\rm out}_{i \kvv} &= \frac{2 \pi}{\hbar} \sum\limits_{\qvv,l} M^{i l}_{i l} \deltoo{i}{l}  (1-\n{l}{\kvv+\qvv}).\qquad
\ea

\end{document}